\documentstyle[prl,epsf,aps]{revtex}

 \newcommand \be {\begin{equation}}
\newcommand \bea {\begin{eqnarray}}
\newcommand \ee {\end{equation}}
\newcommand \eea {\end{eqnarray}}
 \newcommand \BEQ {\begin{equation}}
\newcommand \BEA {\begin{eqnarray}}
\newcommand \EEQ {\end{equation}}
\newcommand \EEA {\end{eqnarray}}
\renewcommand{\d}{{\rm d}}
 \newcommand \eps {\epsilon}
 \newcommand \bi {\bibitem}
\newcommand \s {\sigma}

\newcommand \la {\lambda}
\newcommand \La {\Lambda}

\newcommand \Tr {\mbox{Tr}}
\def\vu{\vspace{1.cm}} 
\def\vh{\vspace{0.5cm}} 
\newcommand \nn {\nonumber\\}
\input{epsf}
\begin{document}
\twocolumn[\hsize\textwidth\columnwidth\hsize\csname@twocolumnfalse\endcsname
\draft \title{Quantum phase transition in spin glasses with multi-spin
interactions}

\author{Theo M. Nieuwenhuizen and Felix Ritort}
\address{Van der Waals-Zeeeman Laboratorium and Institute for
Theoretical Physics\\ University of Amsterdam\\ Valckenierstraat 65\\
1018 XE Amsterdam (The Netherlands).\\ E-Mail:
nieuwenh@phys.uva.nl,ritort@phys.uva.nl}

\date{\today}
\maketitle

\begin{abstract}
We examine the phase diagram of the $p$-interaction spin glass model in
a transverse field. We consider a spherical version of the model and
compare with results obtained in the Ising case. The analysis of the
spherical model, with and without quantization, reveals a phase diagram
very similar to that obtained in the Ising case. In particular, using
the static approximation, reentrance is observed at low temperatures in
both the quantum spherical and Ising models.  This is an artifact of the
approximation and disappears when the imaginary time dependence of the
order parameter is taken into account.  The resulting phase diagram is
checked by accurate numerical investigation of the phase boundaries.
\end{abstract} 

\vfill

\vfill
\vskip.5pc] 
\narrowtext
\section{Introduction}
The interplay between thermal and quantum effects in condensed matter
physics is a longly debated problem \cite{QUANTUM_REVIEWS,BOOK}. The
main differences between both type of effects relies on their dissipative
nature. Thermal physics is inherently dissipative and energy is not
conserved while quantum physics is governed by the Schr\"odinger
equation where energy is conserved if the Hamiltonian does not depend on
time. How to include relaxational effects in a systematic way in the
regime where quantum effects are dominant is a very interesting
open problem \cite{YOSA}.

This question is of the most relevance concerning glassy systems (for
instance structural glasses or spin glasses) which are manifestly non
equilibrium phenomena. Recent developments in the understanding of the
connections between real glasses and spin glasses \cite{BERNA1,BERNA2}
suggest that it is of interest to investigate that family of glassy
models where the static phase transition is continuous from a
thermodynamic point of view (i.e. there is no latent heat) but the order
parameter is discontinuous at the transition temperature. These models
are characterized by a one-step replica symmetry breaking (1RSB)
solution at low temperatures \cite{BOOKS} and the existence of a dynamic
singularity reminiscent of a spinodal instability \cite{KITHWO}. Let us
summarize here the glassy scenario for this type of mean-field
models. At a certain temperature (to be called $T_A$) in the
paramagnetic regime the phase space splits up into different components
or metastable states separated by high energy barriers (divergent with
the size of the system), hence they have infinite lifetime in the
thermodynamic limit. The number of these components is exponentially
large with the size of the system ${\cal N}_s=\exp(S_c)$ where $S_c$ is
the configurational entropy or complexity. From a thermodynamic point of
view the appearance of a large number of states does not induce a
thermodynamic phase transition at $T_A$. Only at a ``Kauzmann''
temperature $T_K$ lower than $T_A$ a true thermodynamic phase transition
(with replica symmetry breaking) is observed. At $T_K$ the complexity
$S_c$ vanishes. Hence, the glass transition is driven by a collapse of
the complexity (entropy crisis) \cite{KITHWO,THEO1}. This is the
mean-field version of the Gibbs-DiMarzio scenario \cite{AGM} for the
glass transition.  The dynamical behavior of the system in the region
$T_K<T<T_A$ is then dominated by the existence of a large number of
components which trap the system for exponentially long time scales
($\tau\sim e^{\alpha N}$ where $N$ is the system size). Whether a sharp
$T_K$ exists in finite dimensions is still a largely unsolved problem
(for recent numerical simulations see \cite{ALFRRI}). 

In this direction, it has been recently proposed a
thermodynamic picture of cooling experiments in spherical $p$-spin
models \cite{NThermo}. This new thermodynamic approach gives an
explanation for the paradox of the Ehrenfest relations at the glass
transition. The main new point in this approach is that
the configurational entropy changes along the transition line.

If the glass transition is driven by a collapse of the configurational
entropy it is natural to ask how this scenario is modified in the
presence of quantum fluctuations. Generally speaking, quantum phase
transitions appear when an external perturbation reaches a
critical value at zero temperature. Because at zero temperature
the entropy vanishes at any value of the external field it is expected
that also the complexity should vanish everywhere at zero temperature
(at least if there is no ground state degeneracy, and this is the 
situation for the mean-field models we will consider here). In the absence of
complexity it is natural to suppose that any adiabatic process at zero
temperature (for instance, a process in which the external field is
slowly turned off) could take the system to the ground state of the
system. If complexity were not fully removed at zero temperature such
expectation would fail since at zero temperature quantum tunneling
processes could not take the system out of the traps during any
adiabatic process, mainly because the height of the barriers is
extremely large \cite{COM1}.

A hint to this problem was recently reported in \cite{RI} where it was
shown that in a certain class of mean-field models where the
Gibbs-DiMarzio scenario is valid, like the random orthogonal model
\cite{ROM}, the complexity vanishes at zero temperature. The transition
turned out to be second order at zero temperature. That proof was
obtained in the framework of the static approximation introduced by Bray
and Moore \cite{BM}. How much general is this result beyond the static
approximation and in other family of models (for instance, the quantum
Potts model \cite{SE}) is still unclear. 

A family of models which has received considerable attention during
the past years are the spherical \cite{CRSO} and Ising \cite{GAR}
$p$-interactions spin-glass models. These models are also characterized
by a classical continuous thermodynamic transition with a discontinuous
jump in the order parameter. The purpose of this work is the study of
the quantum phase transition in this family of models in an external
transverse field. The Ising case has been already considered in the
literature and has revealed novel properties in the phase diagram. In
particular, Goldschmidt \cite{GOL} computed the phase diagram in the
$p\to\infty$ case, i.e. the random energy model of Derrida (hereafter
referred as REM \cite{REM}) in a transverse field. In this case the
static approximation is exact and computations can be easily carried
out. Goldschmidt found a phase diagram with three different thermodynamic
phases, two of them are paramagnetic and separated by a first order
thermodynamic phase transition with latent heat. The existence of first
order phase transitions in spin-glass models with a discontinuous
transition has to be traced back to Mottishaw who studied the random
energy model ($REM$) in an external anisotropy field
\cite{MOT}. Computations for finite $p$ were done later on by Thirumalai
and Dobrosavljevic \cite{DOTH}. They found that the thermodynamic first
order transition line ended in a critical point. Such a critical point
is pushed up to infinite temperature in the $p\to\infty$ limit.
Thirumalai and Dobrosavljevic went further and computed corrections to
the static approximation finding similar qualitative results
at high temperatures. Such an investigation has been recently extended
by De Cesare et al. \cite{CEWARAWA} to the low $T$ region. Corrections
to the $p\to\infty$ limit are generally complicated specially in the
$\beta\to\infty$ limit where the two limits have to be taken in the
appropriate way. In a similar context, recent results by Franz and Parisi
\cite{FRPA} also show the existence of a first order line when two
replicas are coupled in the $T-\epsilon$ plane where $\epsilon$ is the
strength of the coupling between the replica's.

Some computations done in disordered quantum phase transitions involve
the static approximation (hereafter referred as SA) introduced by Bray
and Moore \cite{BM}. This is a reasonable approximation close to the
classical transition line (in particular it predicts a decrease of the
transition temperature as the external field is switched on) but turns
out to be inaccurate at low temperatures where dynamical correlations in
imaginary time start to play a role (this is the reason why the
approximation is called static). To clarify better the physical meaning
of the SA we present an alternative derivation of the mean-field
equations by introducing a solvable spherical version of the quantum
Ising model. We will show that the same scenario for the quantum
transition is valid in both the quantum spherical and quantum Ising
models in the SA as well as beyond it.

For pedagogical reasons we will analyze in detail first the
spherical version of the classical model which is much simpler to
solve. After that we consider a quantum version of the spherical spin
system (recently introduced in~\cite{Nqsm}). We give all the
details that occur in the definition and evaluation of the 
coherent state path integral. After considering a few toy examples
we analyze the quantum spherical $p$-spin model.

Next we will consider the Ising case and estimate the
corrections to the SA numerically solving the self-consistent mean-field
equations.  We will show that the approximation gives a reasonable
estimate (within $10\%$) of the position of the line boundaries which
progressively improves as $p$ increases. Deficiencies of the SA for both
models will be also identified at low enough temperatures, in particular
reentrance of the $T-\Gamma$ boundary line is observed.

The paper is organized as follows. In section II we introduce the
$p$-spin spherical spin glass model and a derivation of the
thermodynamic behavior in a transverse field is obtained with classical
and quantum spins. Section III presents the
solution of the Ising case, the analysis in the SA and also beyond
it. Section V presents the conclusions. Finally some appendices are
devoted to several technical points.

\section{Spherical spins}

In this section we will consider multi-spin interaction
spherical models without and with quantization.  The spherical
model is defined by

\begin{equation}\label{eqt1}
{\cal H}=-\sum_{i_1<i_2<\cdots<i_p} J_{i_1 i_2\cdots i_p}
S_{i_1}^zS_{i_2}^z\cdots S_{i_p}^z-\Gamma\sum_iS^x_i
\end{equation}

where $\Gamma$ is the
the transverse field. The indices $i_1,i_2,\cdots i_p$ 
run from 1 to $N$ where $N$ is the number of sites. 
The $J_{i_1i_2\cdots i_p}$ are couplings Gaussian
distributed with zero mean and variance $p!J^2/(2N^{p-1})$.
The spins have $m$ components and are subject to the spherical
condition 

\be \label{sphconstr}
\sum_{i=1}^N\sum_{a=1}^m S_i^{a\,2}=Nm\sigma
\label{eqt2}
\ee

where $\s$ is a given constant. 

In what follows we first consider the classical case.

\subsection{Classical situation}

The replica calculation for the classical
spherical model without transverse 
field was described by Crisanti and Sommers (CS) \cite{CRSO}. 
The steps are straightforward: 
1) consider $Z^n$; 2) average it over disorder; 3)
rewrite it terms of $q_{\alpha\beta}=(1/N)\sum_{i=1}^N
S^{z}_{i\alpha} S^{z}_{i\beta}$; 4) insert factors 
\bea
1&=&\int_{-\infty}^\infty \d q_{\alpha\beta}
\delta (q_{\alpha\beta}-\frac{1}{N}\sum_{i=1}^N
S^{z}_{i\alpha}S^{z}_{i\beta})
\nonumber\\
&=&\int_{-\infty}^\infty \d q_{\alpha\beta}
\int_{-i\infty}^{i\infty} \frac{N\d \hat q_{\alpha\beta}}{4\pi i}
e^{ \frac{1}{2} \hat q_{\alpha\beta}(N q_{\alpha\beta}-\sum_i
S^{z}_{i\alpha}
S^{z}_{i\beta})}
\eea
A similar representation of the spherical constraints introduces
as Lagrange multipliers the ``chemical potentials'' $\mu_\alpha$.
After these steps, one interchanges the order of integrals.
The remaining integrals over $S^a_{i\,\alpha}$
are all Gaussian (this is the benefit of the spherical approximation)
and  can be integrated out. One is left with an integral over
$q_{\alpha\beta}$, $\hat q_{\alpha\beta}$ and $\mu$, which can be
taken at its saddle point. As for $\Gamma=0$ 
the fields $\hat q_{\alpha\beta}$ can be 
integrated out~\cite{CRSO}, and one ends up with the replicated 
free energy
\begin{eqnarray}
2\beta F_n&=&-\log(\overline{Z_J^n})=-\frac{\beta^2 J^2}{2}\sum_{\alpha\beta}q_{\alpha\beta}^p
-{\rm tr}\ln q \\
&+&\sum_\alpha \left\{\beta\mu_\alpha(q_{\alpha\alpha}-m\sigma)
-\frac{\beta\Gamma^2}{\mu_\alpha}+(m-1)\ln(\beta \mu_\alpha)\right\}
  \nonumber \end{eqnarray}
As for the case $\Gamma=0$ we assume a one-step replica 
symmetry breaking pattern. This involves parameters $\mu$,
$q_d$, $q$, and $x$. These are the chemical potential ($\mu_\alpha=\mu$),
the replica self-overlap ($q_{\alpha\alpha}=q_d$), the
overlap between different replicas inside diagonal 1RSB blocks 
($q_{\alpha\beta}=q$ for $(\alpha,\beta)$ inside a block, while 
vanishing outside the $x\times x$ blocks) 
and the breaking para-mater in the Parisi scheme ($x$ is size of block), 
respectively.
Note that $q_d$ are less than $m\sigma$ since the spins can turn
perpendicular to the $z$-axis. Following CS we obtain for $n\to 0$ 
the ``classical'' free energy $F_{cl}=F_n/n$
\begin{eqnarray}
2\beta F_{classic}=-\frac{\beta^2 J^2}{2}(q_d^p-\xi q^p)
-\frac{1}{x}\ln(q_d-\xi q)-1\nonumber\\ \frac{\xi}{x}\ln(q_d-q)
+\beta\mu(q_d-m\sigma)-\frac{\beta\Gamma^2}{\mu}+
(m-1)\ln(\beta \mu)
\label{bFCS}\end{eqnarray}
where $\xi=1-x$.  Optimization with respect to 
$\mu$, $q_d$, $q$, and  $x$ yields the saddle point relations
\begin{equation}\label{sigma=}
q_d+\frac{\Gamma^2}{\mu^2}+\frac{(m-1)T}{\mu}=m\sigma
\end{equation}
\begin{equation}\label{mu=}
\mu=\frac{p\beta J^2}{2}(q_d^{p-1}-q^{p-1})+\frac{T}{q_d-q}
\end{equation} 

\begin{equation}\label{dq=}
\frac{p\beta^2}{2}q^{p-1}=\frac{q}{(q_d-q)(q_d-\xi q)}
\end{equation} 

\begin{equation}\label{dx=}
-\frac{\beta^2}{2}q^p+\frac{1}{x^2}\ln{\frac{q_d-\xi q}{q_d-q}}
-\frac{q}{x(q_d-\xi q)}=0
\end{equation} 

The latter equation expresses that $\partial F/\partial x=0$.
This means that we consider thermodynamic equilibrium.
For a discussion of the thermodynamics of slow cooling experiments,
see ~\cite{NThermo}.

\subsection{The paramagnet and its pre-freezing line}

Let us first consider the paramagnet, where $q=0$.  Like in the case of
the $p$-spin Ising glass in a transverse field (see the next section),
we find a first order transition line
separating two paramagnetic phases.  This line
is comparable with the boiling line of a liquid and has a critical endpoint. 
 To find it we insert
(\ref{mu=}) with $q=0$ in eq. (\ref{sigma=}) and obtain 
\bea
\Gamma^2=(m\sigma-q_d)(\frac{p\beta J^2}{2}q_d^{p-1}
+\frac{T}{q_d})^2\nonumber\\
-(m-1)(\frac{p J^2}{2}q_d^{p-1}+\frac{T^2}{q_d}) \eea
At large $T$ and $\Gamma$ this has just one real positive 
solution $0<q_d<m\sigma$.
However, below a critical value $T_{cep}$ there is
a regime of $\Gamma$-values where there occur three
solutions rather than one.  The outer ones are stable, while the middle
one is unstable.  This critical endpoint ($cep$) has coordinates
$(T_{cep},\,\Gamma_{cep})$, determined by $\d\Gamma/\d q_d=\d^2\Gamma/\d
q_d^2=0$. From this point a first order transition line
originates towards the spin glass phase and intersects it at the
multi-critical point $(T_{mcp},\Gamma_{mcp})$ \cite{p=3cr}.
Along this line there is a finite latent heat, that vanishes at
the critical end point.
 It separates a small transverse field
phase with large  ordering in the $z$-direction $(q_d^>)$ 
from a phase with smaller ($q_d^<$) ordering in the $z$-direction
on the large field side. 

In analogy with wetting phenomena, where a pre-wetting line occurs
off coexistence, we call this {\it the pre-freezing line}.
In order to motivate this term, let us first explain the 
situation of first order wetting of a bulk fluid A by 
a thin layer of a fluid B  ~\cite{PSW,FLN}.
At bulk coexistence of A and B phases there
is a wetting temperature $T_w$. For $T$  below  $T_w$ a finite layer
(``wetting layer'')
of B atoms will cover the A phase; for first order wetting 
this layer remains finite in the limit $T\to T_w^-$. 
For $T>T_w$ there will be an infinite
B layer (``complete wetting''). When the fluids A and B are off coexistence
there is a difference in chemical potential $\Delta\mu$. Let us take the
convention that on the $\Delta\mu<0$-side the B layer is always finite.
Then when $\Delta\mu\to 0^-$ for $T<T_w$, the B layer will reach its finite
thickness discussed for $\Delta\mu=0$. For $\Delta\mu>0^+$ it will be infinite
however, leading to a discontinuous transition. For $T>T_w$, however,
the thickness of the B-layer will diverge continuously in the limit
$\Delta\mu\to 0^-$, in order to be infinite at $\Delta\mu=0$, and it will
remain infinite for $\Delta \mu>0$. In that
temperature regime there is a continuous transition at $\Delta\mu=0$.
Thermodynamics requires coexistence at a first order
transition line (called the ``pre-wetting line'') which separates the
regime of continuous and first order wetting. The endpoint of this line
is called the ``pre-wetting critical point''. 
The pre-wetting line, occurrence of hysteresis,
and, near the pre-wetting critical point,
scaling of the  jump in coverage across the line 
have been observed in $^4$He on Ce ~\cite{RutTab} and for 
methanol-cyclohexane mixtures~\cite{Bonn}~\cite{Kellay}.

In our spin glass a similar situation occurs. The SG-PM$^>$ line is a
 continuous  transition line (in the sense that there is no latent
heat), whereas we will find a finite latent heat at the SG-PM$^<$
transition. Also in this situation a first order line with non-vanishing
latent heat  must emerge from the point $(T_{mcp},\Gamma_{mcp})$ 
and divide the paramagnet into two regimes. 
It is the line discussed, and by analogy we propose to call it the
pre-freezing line. Its critical endpoint can then be called the 
``pre-freezing critical point''.

\subsection{The spin glass phase}
This discontinuity of the paramagnet has no analog in the spin glass.
There is only one spin glass phase, namely the continuation of 
the $q_d^>$ paramagnet, with continuous $q_d$ at the transition line $x=1$.
 We stress that this also holds also when the PM$^<$ phase is the 
thermodynamically stable phase: also then the (metastable) SG phase
merges with the (metastable) PM$^>$ phase at the $x=1$ line.

To check this continuity of the SG-phase, let us insert
eq. (\ref{dq=}) into eq. (\ref{dx=}) and replace the $x$-dependence by
dependence on a new variable $\eta$ via 
\be\label{x=}
x=\frac{p-1-\eta}{\eta}\,\,\frac{q_d-q}{q} 
\ee 
Eq. (\ref{dx=}) then becomes 
\be \ln\frac{p-1}{\eta}=\frac{(p-1-\eta)(\eta+1)}{p \eta} \label{eqeta}\ee
which has a solution $0<\eta<1$.  This shows that $\eta$ is independent
of $\Gamma$ and $T$.  (For $\Gamma=0$ this was noted already by Crisanti
and Sommers~\cite{CRSO}).  Once $\eta$ is known, we can choose $x$ and solve
$q=(p-1-\eta)q_d/(p-1-\xi\eta)$ from (\ref{x=}).  
Eq. (\ref{dq=}) will then yield 
\be
q_d=\frac{p-1-\xi\eta}{p-1-\eta}\left(\frac{2T^2(p-1-\eta)^2} {x^2\eta
p(p-1)}\right)^{1/p} \ee 
At fixed $x$ we can vary $T$. We thus know
$q_d$ and $q$, and therefore find a curve $\Gamma(T)$.  At small enough
$T$ two values of $x$ can lead to a given point ($T,\Gamma$); we need
the smallest of these two $x$-values.  By varying $x$ between $1$ and $0$
this procedure then uniquely
determines the spin glass phase.

The pre-freezing line intersects the PM$^>$-SG transition line at a 
multi-critical point $(T_{mcp},\Gamma_{mcp})$~\cite{p=3cr}. 

Just as at $\Gamma=0$ the transition PM$^>$-SG occurs with $x=1$. This
is a thermodynamic continuous phase transition. The SG free energy
exceeds the PM$^>$ one by an amount of order $\xi^2$.  The transition
PM$^<$-SG is thermodynamically first order and occurs with $x<1$.  The
transition line is fixed by equating the free energies of the PM$^<$ and
the SG solutions.

\subsection{Low temperature behavior}
The transition line between PM$^<$ and SG will continue down to $T=0$.
Everywhere along this line there will be a latent heat accorded 
by a jump in entropy. For studying the low $T$ behavior we set
\be
q_d=(1+Tr)q
\ee
which implies
\be
x=\frac{p-1-\eta}{\eta}rT;\qquad \frac{p(p-1)J^2}{2\eta}q^pr^2=1
\ee
\be
\mu=\frac{\eta}{qr}\,\,\frac{(1+rT)^{p-1}-1}{(p-1)rT}
+\frac{1}{qr}
\ee

At $T=0$ this yields $\mu=(\eta+1)/qr$ and then
\be
q+\frac{2\eta \Gamma^2}{p(p-1)(\eta+1)^2J^2}q^{2-p}=m\sigma
\ee
For $p=3$ it can be solved exactly
\be
q=\frac{m\sigma}{2}+\sqrt{\frac{m^2\sigma^2}{4}-
\frac{\eta\Gamma^2}{3(\eta+1)^2J^2}}
\ee
showing that the SG phase cannot exist at large $\Gamma$.
For small $\Gamma$ the spin glass phase is stable; for larger
values it becomes meta-stable and for still larger values it
will be unstable.
The free energy may be expanded in powers of $T$
\be \label{FsmallT}
F=F_0+T(-\frac{m}{2}\ln T +F_1)+{\cal O}(T^2)
\ee
One finds
\bea
F_0&=&-\frac{\eta+2}{2pr}-\frac{\Gamma^2qr}{\eta+1}\nonumber\\
F_1&=&\frac{1}{2}(\eta+1+(m-1)\ln(\eta+1)+m\ln qr)
\eea

In the paramagnet PM$^<$ one has $q_d\approx
T\sqrt{m\sigma}/\Gamma$, 
$\mu\approx \Gamma/\sqrt{m\sigma}$, implying
\bea
F_0&=&-\Gamma\sqrt{m\sigma}\\
F_1&=&\frac{1}{2}(1+m\ln \frac{\Gamma}{\sqrt{m\sigma}})
\eea
Equating the $T=0$ results we find a transition at some $\Gamma_c$.
 For small $T$ we get
\be 
F_{\rm SG}-F_{\rm PM}=A(\Gamma-\Gamma_c)+BT
\ee
with $A>0$  because the stable phase has lowest free energy, and
\be
 B=\frac{1}{2}(\eta-\ln(\eta+1))+\frac{m}{4}\ln\frac{m\sigma}
{m\sigma-q}
\ee
also positive. For small $T$ the transition line has a linear slope
\be
\Gamma=\Gamma_c-\frac{B}{A} T
\ee
showing that there occurs no reentrance.
For the case $m=2$ and in units where $m\sigma=1$,
the full phase diagram for $p=4$ is shown in figure 1. In figure 2
we show the latent heat as a function of the temperature. It has been
computed along the boundary lines starting from
the critical point, following the PM$^<$-PM$^>$ and
the PM$^<$-SG lines. Note the existence of a sharp maximum at
$T=T_{mcp}$. 

\begin{figure}
\centerline{\epsfxsize=8cm\epsffile{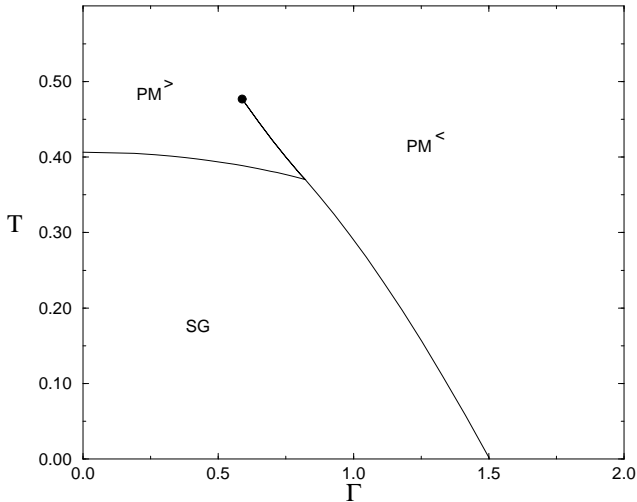}}

\caption{Phase diagram for the classical spherical model for $p=4$,
$m\sigma=1$, $m=2$. 
The multi-critical point and the critical 
point are given by $T_{mcp}=0.3703, \Gamma_{mcp}=0.8208,
T_{cep}=0.4767,\Gamma_{cep}=0.5878$ while at $T=0$, $\Gamma_c=1.503$}
\end{figure}

\begin{figure}
\centerline{\epsfxsize=8cm\epsffile{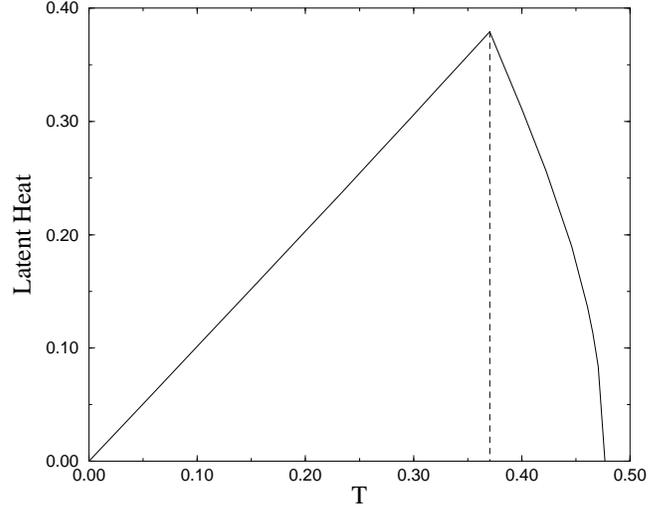}}

\caption{Latent heat for the classical spherical model for $p=4$ versus
$T$ along the boundary line which separates the PM$^<$ phase from the
PM$^>$ (right part above the multi-critical point) and SG phases (left
part below the multi-critical point). There is a maximum at the
multi-critical point (indicated by the dashed line).}
\end{figure}

\subsection{Quantum spherical spins}
Due to the form (\ref{FsmallT}) the entropy of spherical models diverges 
as $(m/2)\ln T$ for small $T$.  The related zero temperature 
specific heat $C=m/2$, occurring due to the Gaussian nature of the spins,
is  analogous to the Dulong-Petit law of classical harmonic oscillators.  
In order to have a physical description in the low $T$ regime, one of us
recently proposed quantization by analogy with harmonic oscillators 
\cite{Nqsm}. Here we present some details of this approach.
It follows the standard Trotter-Suzuki approach of thermal field theories,
see e.g. the book of Negele and Orland~\cite{NegeleOrland}.

The approach starts from the Trotter formula of the path integral
representation of the partition sum, in which the coherent state
representation of the identity is restricted to coherent states
described by parameters which satisfy the spherical constraint. For a
set of harmonic oscillators ${\bf S}=\{S_i^a\}$ with $(i=1,\cdots,N$,
$a=1,\cdots,m)$ a coherent state is defined by \be |{\bf S}\rangle
=e^{{\bf S\cdot S}_{op}^\dagger}|{\bf 0}>
=\prod_{i,a}\left\{\sum_{n_{ia}=0}^\infty
\frac{(S_i^a)^{n_{ia}}}{\sqrt{n_{ia}!}}|n_{ia}\rangle \right\}\ee where
$S_{i\,op}^{a\,\dagger}$ is the creation operator of the harmonic
oscillator $(i,a)$ and $S_i^a$ is a $c$-number.  The coherent states are
overcomplete and have innerproduct \be \label{chsip} \langle{\bf
S}'|{\bf S}\rangle =e^{ {\bf S}'\cdot {\bf S}} \ee which can be checked
in various ways.  The coherent state representation of the identity in
Fock space reads in general \bea {\bf 1}&=&
\prod_{i,a}\left\{\sum_{n_{ia}=0}^\infty |n_{ia}\rangle \langle
n_{ia}|\right\}\nonumber\\ &=&\int \prod_{ia}\frac{\d S_i^{a\,\ast} \d
S_i^a}{\pi} e^{-{\bf S^\ast\cdot S}}|{\bf S}\rangle \langle {\bf S}|
\eea It was proposed by one of us ~\cite{Nqsm} to enforce the spherical
constraint by restricting this representation to coherent states which
satisfy the spherical constraint \be {\bf S^\ast\cdot
S}=\sum_{i=1}^N\sum_{a=1}^m S_i^{a\ast}S_i^a=Nm\sigma \ee This is done
by replacing \bea \label{spherid} {\bf 1}&\to&{\bf 1}_{spherical}\\
&\equiv& C\int \prod_{ia}\frac{\d S_i^{a\,\ast} \d S_i^a}{\pi} e^{-{\bf
S^\ast\cdot S}} |{\bf S}\rangle \langle {\bf S}| \delta({\bf S^\ast\cdot
S}-Nm\sigma)\nonumber \eea where the constant $C$ will be fixed later.
In a Trotter approach one calculates \be Z={\rm tr}\,\, e^{-\beta H({\bf
S}_{op}^\dagger,{\bf S}_{op})}= {\rm tr} \left( e^{-\epsilon H({\bf
S}_{op}^\dagger,{\bf S}_{op})}\right) ^M \ee with
$\epsilon=\beta/M$. Between all factors $\exp(-\epsilon H)$ one inserts
the coherent state representation of the identity. For our spherical
spins we have to choose the truncated versions (\ref{spherid}).  Let us
number them by $j=1,\cdots,M$.  It was shown by Negele and Orland that
for normal ordered Hamiltonians \be\label{matrixel} \langle{\bf S}_{j}|
e^{-\epsilon H({\bf S}_{op}^\dagger,{\bf S}_{op})} |{\bf
S}_{j-1}\rangle=e^{{\bf S}_{j}^\ast\cdot{\bf S}_{j-1}- \epsilon H({\bf
S}_{j}^\ast,{\bf S}_{j-1})} \ee The term $\exp({\bf S}_j^\ast\cdot{\bf
S}_{j-1})$ arises from the overlap of the coherent states (\ref{chsip}),
while the $\epsilon H$ correction can be found by expanding the
exponential, using ${\bf S}_{op}|{\bf S}_{j-1}\rangle={\bf S}_{j-1}|{\bf
S}_{j-1}\rangle$ and its Hermitean conjugate $\langle{\bf S}_j|{\bf
S}^\dagger_{op}=\langle{\bf S}_j|{\bf S}_j^\ast$, and re-exponentiating
the result.  Corrections are of order $\epsilon^2$ and can be neglected
in the limit $M\to\infty$ (for a discussion, see ~\cite{NegeleOrland}).
Introducing the imaginary time variable $\tau=j\epsilon=j\beta/M$ and
writing out the spherical constraints in terms of an imaginary valued
chemical potential $\mu(\tau)$, this leads to the coherent state path
integral representation or thermal field theory for spherical spins \bea
Z&=&\int D\mu D{\bf S}^\ast D{\bf S}\exp(-A)\label{eq33}\eea with
integration measure \bea \int D{\bf S}^\ast D{\bf S}&=&\prod_{ia\tau}
\int_{-\infty}^\infty\int_{-\infty}^\infty \frac{\d \Im(S_{i}^a(\tau))\d
\Re (S_{i}^a(\tau))}{\pi}\nn \int D\mu &=&C_M\prod_\tau
\int_{-i\infty}^{i\infty} \frac{\epsilon\d\mu(\tau)}{2\pi i} \eea
involving the constant $C_M=C^M$ to be fixed below and the action \bea
A=\sum_\tau &\d\tau&\{ {\bf S}^\ast(\tau)\cdot \frac{\d {\bf
S}(\tau)}{\d \tau}+ \mu(\tau)({\bf S}^\ast(\tau)\cdot{\bf
S}(\tau)-Nm\sigma) \nonumber\\ &+& H({\bf S}^\ast(\tau),{\bf
S}(\tau-\d\tau))\} \eea where $\d\tau\equiv\epsilon$ and $\d {\bf
S}(\tau)/\d\tau\equiv ({\bf S}(\tau)-{\bf S}(\tau-\d\tau)/\d\tau$
involves ${\bf S}(\tau)$ due to eq. (\ref{spherid}) and ${\bf
S}(\tau-\d\tau)$ due to eq. (\ref{matrixel}).  The trace structure leads
to periodic boundary conditions ${\bf S}(\beta)={\bf S}(0)$, related to
the bosonic nature of the spherical spins. One might be tempted to take
the continuum limit of eq.(\ref{eq33}). However, note that this is a
dangerous limit since problems can arise that do not occur in our
discrete formulation \cite{NegeleOrland}. For a concrete example we have
discussed in Appendix A what is the origin of the problem.

\subsubsection{Free spins in a field}

The simple case of free spherical spins in an external field is 
already non-trivial~\cite{Nqsm}. This is because the spherical
constraint couples the spins.
Let us consider the Hamiltonian
\bea 
H = -\Gamma \sum_i ( S^{x\,\dagger}_{i\,op}+S_{i\,op}^x) \eea
We can introduce imaginary time Fourier transforms
\bea
{\bf S}_i(\tau)&=&\sum_{\omega} {\bf S}_{i \omega}
e^{-i\omega \tau} \nonumber\\
{\bf S}_{i \omega}&=&\frac{1}{M}\sum_\tau 
{\bf S}_i(\tau)e^{i\omega \tau}
\eea
where $\omega=2\pi nT$ is a Matsubara frequency 
with $1\le n\le M$ and $\tau=j\beta/M$ with 
$1\le j\le M$.
Integrating out the spins we obtain 
\bea
Z=\int D\mu\exp(-NA)
\eea
with (denoting imaginary times again by $j=\tau/\epsilon$), 
\bea
A=-\epsilon m\sigma\sum_j \mu(j\epsilon)+m\,\,{\rm tr}_j
\ln B-{\epsilon^2\Gamma^2}\sum_{jj'}B^{-1}_{jj'}
\eea
where
\bea 
B_{jj'}=(1+\epsilon\mu(j\epsilon))\delta_{jj'}-
\delta_{j,j'+1}
\eea
with $\delta_{1,M+1}\equiv 1$ due to the periodic boundary condition.
We write 
\be \mu(j\epsilon)= \mu+ \mu_j(1+\epsilon\mu)\ee 
where $\mu$ is the saddle point value, that will turn out to be
real, whereas the deviations $\mu_j$ are imaginary and turn out
to be ${\cal O}(N^{-1/2})$.
We expand to second order in $\mu_j$. The matrix
\be \bar B_{jj'}=\frac{1}{1+\epsilon\mu}B_{jj'}(\mu)\ee
has diagonal elements 1 and off-diagonal elements 
$-a=-1/(1+\epsilon\mu)$. 
Its inverse is ~\cite{NegeleOrland}
\bea \bar B_{ij}^{-1}&=&\frac{1}{1-a_\lambda ^M}\,\,
a_\lambda ^{i-j} \qquad i\ge j 
\nonumber\\
&=&\frac{1}{1-a_\lambda ^M}\,\,a_\lambda ^{M+i-j} \qquad i < j
\eea
We can now expand the action to second order in $\mu_j$. This gives 
after some algebra
\be\label{a012} A=A_0+A_1+A_2\ee  
$A_0$ is the saddle point free energy
\bea
\beta F=A_0&=&-\beta\mu  m\sigma+m\ln[(1+\epsilon\mu)^M-1]
-\frac{\beta\Gamma^2}
{\mu } \nn
&\to& -\beta\mu m\sigma+m\ln[e^{\beta\mu}-1]
-\frac{\beta\Gamma^2}{\bar \mu}
\eea
This gives the saddle point equation
\BEQ
\frac{m}{1-e^{-\beta\mu}}+\frac{\Gamma^2}{\mu^2}=m \sigma
\EEQ
At zero field one has $e^{\beta\mu}=\sigma/(\sigma-1)$, yielding
$\beta F=-S_{\infty}$ with infinite temperature entropy
\BEQ S_{\infty}=m[\sigma\ln\sigma-(\sigma-1)\ln(\sigma-1)] \EEQ
Due to the scaling of the spherical constraint with $m$
and the harmonic nature of the spherical spins, this yields an
equal amount of entropy  for each spin direction.

For non-zero field the large temperature behavior is still of this
form. For small temperatures, however, excitations will have a gap
$\Delta E=\mu(T=0)$$=$$\Gamma/\sqrt{m(\sigma-1)}$. This follows since,

\BEQ
\mu=\frac{\Gamma}{\sqrt{m(\sigma-\frac{e^{\beta\mu}}{e^{\beta\mu}-1})}}
\approx \frac{\Gamma}{\sqrt{m(\sigma-1)}}
(1+\frac{e^{-\beta\Delta E}}{2(\sigma-1)})~~~.
\EEQ

Note that this gap scales linearly in the field $\Gamma$, as expected
for free spins in a field. Other quantization schemes have been proposed
where the action involved a second oder derivative in imaginary time
\cite{Hart-Weichman},\cite{Voita}, \cite{VoitaSchreiber}.  Physically
this is due to a kinetic term of the form $(\d S/\d\tau)^2$ rather than
our first order derivative $S^\ast\d S/\d\tau$ arising from the Trotter
approach. The kinetic terms describe different physics, e.g. the
quantized kinetic energy of a rotor. Such a system always has a finite
energy gap due to its harmonic oscillator character. Spin systems are
fundamentally different.  Spins have no kinetic energy, and for
quantized spherical spins the energy gap indeed vanishes when the field
vanishes.

The next terms in eq. (\ref{a012}) fix the prefactor of the
path integral. They are discussed in appendix B. 

\subsubsection{Pair couplings} 

Another non-trivial situation is quantized spherical spins that are
coupled in pairs in the presence of an external field. This covers both
the ferromagnet  and the spin glass cases.
\bea 
H =  -\sum_{ij}J_{ij}{\bf S}_{i\,op}^\dagger{\bf S}_{j\,op}
-\sum_i \Gamma_i( S^{x\,\dagger}_{i\,op}+S_{i\,op}^x) \eea
We can diagonalize the coupling matrix, and
introduce imaginary time Fourier transforms
\bea
{\bf S}_i(\tau)&=&\sum_{\omega}\sum_\lambda {\bf S}_{\lambda \omega}
e_i^\lambda e^{-i\omega \tau} \nonumber\\
{\bf S}_{\lambda \omega}&=&\frac{1}{M}\sum_{i \,\tau }
{\bf S}_i(\tau)e^{i\omega \tau}e_i^\lambda 
\eea
where $e_i^\lambda$ is the normalized 
eigenvector of $J_{ij}$ with eigenvalue $J_\lambda$.
Integrating out the spins we obtain the intensive free energy
\bea\label{Fpair} 
\beta F&=&-\beta\mu m\sigma+mM\ln(1+\epsilon\mu)\\
&+&\frac{1}{N}\sum_\lambda(m\ln(1-a_\lambda ^M)
-\frac{M\epsilon^2\Gamma_\lambda^2}
{1-a_\lambda }) \nn
&\to& -\beta\mu m\sigma+m\int\rho(J_\lambda)\d J_\lambda
\ln(e^{\beta\mu}-e^{\beta J_\lambda})\nn
&-&\int\rho(J_\lambda)\d J_\lambda
\frac{\beta\Gamma_\lambda^2}{\mu-J_\lambda}
\eea
where $\Gamma_\lambda=\sum_ie_i^\lambda\Gamma_i$ is the projection
 of the field on eigenstate $\lambda$.
The thermally averaged occupation numbers are
\BEQ
\langle S^\ast_{\lambda\omega}S_{\lambda\omega}\rangle=
\frac{T}{\mu-i\Omega_\omega-J_\lambda e^{i\epsilon\omega}}
\EEQ
These  results have been analyzed for a ferromagnet on a simple cubic
lattice. One has $J_\lambda\to J(k)=2J(\cos k_x+\cos k_y+\cos k_z)$,
with integration measure $\d^3 k/(2\pi)^3$. In particular one finds
a low temperature specific heat $C\sim m T^{3/2}$ 
due to spin waves in $m$ directions. Note that the spherical
constraint also allows longitudinal spin waves~\cite{Nqsm}.

When spins are only
coupled in the $z$-direction, while the field acts in the transverse
($x$) direction, one has in the limit $M\to\infty$ 
\BEA
\beta F&=&-\beta\mu m\sigma+\int\frac{\d^3 k}{(2\pi)^3}
\ln(e^{\beta\mu}-e^{\beta J(k)})\nn
&+&(m-1)\ln(e^{\beta\mu}-1)
-\frac{\beta \Gamma^2}{\mu-6J}
\EEA
From these equations the zero-temperature quantum phase transition
in a transverse field can be analyzed. Some results were given
in ref. \cite{Nqsm}.

The case of a spherical spin glass with pair couplings between the
$z$-components  will be useful
for a check of the results of next section.
Here $J_{ij}$ are random Gaussian with average zero and variance
$J^2/N$. The distribution of eigenvalues is the semi-spherical law
\BEQ
\rho(J_\lambda)=\frac{1}{2\pi J^2}\sqrt{4J^2-J_\lambda^2}
\EEQ
with $-2J<J_\lambda<2J$.
For any $M$ one can now calculate the thermal occupation numbers
\BEA
\hat q_{d\omega}&=&\int \rho(J_\lambda)\d J_\lambda
\langle S^\ast_{\lambda\omega}S_{\lambda\omega}\rangle\nn
&=&
\int \rho(J_\lambda)\d J_\lambda
\frac{T}{\mu-i\Omega_\omega-J_\lambda e^{i\epsilon\omega}}\nn 
&=&\frac{2T}{\mu-i\Omega_\omega+
\sqrt{(\mu-i\Omega_\omega)^2-4J^2 e^{2i\epsilon\omega}}}
\EEA
Below the phase transition the system will have condensed partly
in the mode with largest eigenvalue $2J$. 
Its occupation number $Nq\equiv S^\ast_{2J}S_{2J}$ will be 
extensive. For $\omega=0$ one then has the expectation value
\BEQ\label{qdtilde}
\tilde q\equiv \hat q_{d\,\omega=0}=q+\frac{2T}{\mu+
\sqrt{\mu^2-4J^2}}
\EEQ
The free energy reads
\BEA\label{Fp21}
\beta F&=& -\beta\mu m\sigma+
\int\rho(J_\lambda)\d J_\lambda\nn
&\times&
\sum_\omega\ln(1+\epsilon\mu-
(1+\epsilon J_\lambda)e^{2i\epsilon\omega})
\nn
&+&(m-1)\sum_\omega \ln(1+\epsilon\mu-
e^{2i\epsilon\omega})-\frac{\beta\Gamma^2}{\mu}\nn
&+&\beta(\mu-2J)q
\EEA
where the $(m-1)$-terms arise from the transverse spin components 
and the last term from the ordering field, respectively.
Using
\BEA
&\int& \rho(J_\lambda)\d J_\lambda\ln(a-bJ_\lambda)=\nn
&\ln&\frac{a+\sqrt{a^2-4b^2}}{2}
+\frac{a-\sqrt{a^2-4b^2}}{2(a+\sqrt{a^2-4b^2})}
\EEA
for $a=1+\epsilon\mu-e^{2i\epsilon\omega}
$, $b=\epsilon e^{2i\epsilon\omega}$, we can express
 the integral as
\BEA
-1&+&\beta\mu(\tilde q_{d}-q)
-\ln(M(\tilde q_{d}-q))
-\frac{\beta^2J^2}{2}(\tilde q_{d}-q)^2+\\
\sum_{\omega\neq 0}&[&-1+\beta(\mu-i\Omega_\omega)\hat q_{d\,\omega} 
-\ln(M\hat q_{d\,\omega})
-\frac{\beta^2J^2}{2}e^{2i\epsilon\omega}\hat q_{d\,\omega}^2]
\nonumber
\EEA
 Variation of eq. (\ref{Fp21}) 
wrt $S_{2J}$ yields $q\equiv S_{2J}^\ast S_{2J}/N =0$ 
or $\mu=2J$, $q>0$. In the latter case
eq. (\ref{qdtilde}) yields $\tilde q_d=q+T/J$. 
We may therefore make the replacement
\BEQ
-2\beta J q+\beta J \tilde q_d q-\frac{\beta^2J^2}{2}q^2\to
\frac{\beta^2J^2}{2}q^2-\frac{q}{\tilde q_d-q}
\EEQ
This finally leads to 
\BEA\label{Fp=2}
\beta F&=&-\frac{\beta^2J^2}{2}(\tilde q_d^2-q^2)
-\frac{q}{\tilde q_d-q}-1+\beta\mu(\tilde q_d-m\sigma)\nn
&-&
\ln[\beta\mu(\tilde q_d-q)-\frac{\beta\Gamma^2}{\mu}
+m\ln[(1+\epsilon\mu)^M-1]\nn
&+&\sum_{\omega\neq 0}
[-1+\beta(\mu-i\Omega_\omega)\hat q_{d\,\omega}
-\ln(\beta(\mu-i\Omega_\omega)\hat q_{d\,\omega})\nn
&-&\frac{\beta^2J^2}{2}e^{2i\epsilon\omega}
\hat q_{d\,\omega}^2]
\EEA
In next section we shall recover this expression 
as the $p=2$, $x\to 0$  case of eqs. (\ref{Fst+qu}),
(\ref{Fstatic}), (\ref{Fquant}).
The condition $x=0$ occurs 
due to absence of replica symmetry breaking.
This should be expected since the system condenses in only one mode,  
the one having the largest eigenvalue~\cite{KTJ}.
When also short range ferromagnetic interactions are present,
thermodynamics and correlation functions can be solved exactly.
The largest mode may then  be due to the onset of spin glass
ordering or of ferromagnetism.~\cite{N85}

For low $T$ the representation (\ref{Fpair}) shows that the specific
heat behaves as $T^{3/2}$. One can also  determine the time-dependent
correlation function
\BEA
q_d(\tau)&=&\sum_\omega\hat q_{d\,\omega}e^{i\omega\tau} \nn
=&\int& \rho(J_\lambda)\d J_\lambda\frac{e^{\tau(\mu-\lambda)}}
{1-e^{-\beta(\mu-\lambda)}}\qquad (-\beta<\tau\le 0)
\EEA
At $T=0$ it is unity in $\tau =0$ and zero in $0^+$, while it decays as
$q_d(\tau)\sim |\tau|^{-3/2}$ for $\tau\to-\infty$.

\subsection{Spin glass in a transverse field}

It is known that a given classical Hamiltonian may come from several
quantum Hamiltonians. A similar situation occurs here.
The simplest case is where the classical Hamiltonian contains complex 
valued spins~\cite{Nqsg}, which can be replaced by operators.
We thus consider the case where $H$ depends at each site either
on the creation or the annihilation operator.  
For $p=4$ we have the Hamiltonian 
\bea \label{quH4}
H&=&\frac{-1}{4!}\sum_{i,j,k,l=1}^N J_{ijkl}S^{z\,\dagger}_{i\,op} 
S^{z\,\dagger}_{j\,op} S^z_{k\,op} S^z_{l\, op} \nonumber\\
&-&\Gamma\sum_i( S^{x\,\dagger}_{i\,op}+S_{i\,op}^x) \eea
with Hermitean $J_{ijkl}$ ~\cite{Nqsg}.
(For odd $p$ we have to add Hermitean conjugate terms; here they
are included already.)  This means that 
for each quartet $(i,j,k,l)$ with $i<j$, $k<l$ there are four independent
random variables,
$J_{1,2}'$ and $J_{1,2}''$, each having average zero and
 variance $9J^2/N^3$. In terms of
$J_{1,2}=J_{1,2}'+iJ_{1,2}''$ the couplings in eq. (\ref{quH4})
read $J_{ijkl}=J_{klij}^\ast=J_1+iJ_2$ and
$J_{ilkj}=J_{kjil}^\ast=J_1-iJ_2$. These results hold for $i<j$, $k<l$.
The other sectors follow,
of course, by symmetry:
$J_{ijkl}=J_{jikl}=J_{ijlk}=J_{jilk}$. A similar approach will 
work for general even $p$.

In the replicated free energy
we look for a saddle point with $\mu_\alpha(\tau)=\mu$ and 
$q_{\alpha\alpha}(\tau,\tau')=\langle S^z_{\alpha}(\tau)^{\ast}
S^z_{\alpha}(\tau')\rangle= q_d(\tau-\tau')$ 
and with $q_{\alpha\beta}(\tau,\tau')=\langle S^z_{\alpha}(\tau)^{\ast}
S^z_{\beta}(\tau')\rangle=q_{\alpha\beta}$ independent of
$\tau,\tau'$ for $\alpha\neq\beta$. For $M$
Trotter steps the free energy for a one step replica symmetry breaking
solution with plateau $q$ and breakpoint $x=1-\xi$ reads 
\be\label{Fst+qu}
F=F_{static}+F_{quant} \ee with
\bea\label{Fstatic} &\beta& F_{static}=
-\frac{\beta^2J^2}{2}(\tilde q_d^p-\xi q^p)
-\frac{1}{x}\ln\frac{\tilde q_d-\xi q}{\tilde q_d-q}-1
\\
&+&\beta\mu(\tilde q_d-m\sigma)-\ln[\beta\mu(\tilde q_d-q)]
-\frac{\beta\Gamma^2}{\mu}
+m\ln[e^{\beta\mu}-1]\nonumber\label{fs} \eea 
being mainly twice as large as
$F_{classic}$ in (\ref{bFCS}), due to doubling of spin degrees of
freedom (now complex, previously real). A more important difference
is the replacement $m\ln\beta\mu\to m\ln(e^{\beta\mu}-1)$. As
we shall see, this improves quite a bit on the not-too-low
temperature behavior. 
After deriving this expression at finite $M$ we have replaced a term 
$m\ln[(1+\epsilon \mu)^M-1]$ (see also eq. (\ref{Fp=2}))
by its limit
$m\ln[e^{\beta\mu}-1]$; we shall come back to this point below.
The quantum correction reads at finite $M$
\bea\label{Fquant} &\beta&
F_{quant}\nn
&=&\sum_{\omega\neq 0} [-1+\beta(\mu-i\Omega_\omega)
\hat{q}_{d\,\omega}
-\ln(\beta(\mu-i\Omega_\omega)\hat{q}_{d\,\omega})]
\nonumber\\
&-&\frac{\beta^2 J^2}{2M}\sum_{\tau}
q_d^{p/2}(\epsilon+\tau)q_d^{p/2}(\epsilon-\tau)
+\frac{\beta^2 J^2}{2}\tilde q_d^p\label{fq}
\eea 
Here we have Fourier transforms
\be
q_d(\tau)=\sum_\omega \hat q_{d\,\omega}e^{i\omega \tau}
\qquad
\hat q_{d\,\omega}=\frac{1}{M}\sum_{\tau} q_d(\tau)e^{-i\omega \tau}
\ee
and denoted $\tilde q_d\equiv \hat q_{d\,\omega=0}$.

If one inserts in the $J^2$ term of $F_{quant}$ that $q_d(\tau)=\tilde
 q_d$ is independent of $\tau$, the $J^2$-terms cancel. Then the
 $q_\omega$'s can be solved, after which the whole $F_{quant}$ vanishes
 identically.  This is closely related to the static approximation (SA)
 of Ising models (see below) introduced by Bray and Moore
 \cite{BM}. This approximation neglects the time dependence of the
 correlator $q_d(\tau)$
 where $q_d(\tau)\to q_d$ being also independent of $\tau$. The remaining
 difference with previous classical theory is the replacement
 $m\ln\beta\mu\to m\ln(\exp(\beta\mu)-1)$ in going from $2F_{classic}$
 to $F_{static}$. This replacement already improves the low temperature
 behavior. The phase diagram in the SA can be numerically computed and
 is shown in figure 3. It is qualitatively similar to that computed in
 the classical case (see figure 1). We find a thermodynamically
 first-order transition line with a multi-critical point terminating in
 the high-T phase in a critical end point. At low temperatures the first
 order line shows the phenomena of reentrance and negative latent heat
 along that line down to $T=0$. This is a failure of the SA as we will
 show below.

Beyond the static
 approximation the saddle point equations read:
\begin{eqnarray}\label{sigmaqu=}
&\frac{\partial}{\partial\mu}=0\to&
 q_d(\beta)+\frac{\Gamma^2}{\mu^2}
+(m-1)\frac{e^{\beta\mu}}{e^{\beta\mu}-1}
=m \sigma
\\
\label{muqu=}
&\frac{\partial}{\partial\tilde q_d}=0\to&
\beta\mu=\frac{1}{\tilde q_d-q}-\frac{p\beta^2 J^2}{2} q^{p-1}
\\&\,&+
\frac{p\beta^2 J^2}{2M}\sum_\tau q_d^{p/2-1}(\epsilon-\tau)
q_d^{p/2}(\epsilon+\tau)\nn
\label{dqqu=}
&\frac{\partial}{\partial q}=0\to&\frac{p\beta^2J^2}{2}q^{p-1}
=\frac{q}{(\tilde q_d-q)(\tilde q_d-\xi q)}\\
\label{dxqu=}
&\frac{\partial}{\partial x}=0\to&
-\frac{\beta^2J^2}{2}q^p+\frac{1}{x^2}
\ln{\frac{\tilde q_d-\xi q}{\tilde q_d-q}}
-\frac{q}{x(\tilde q_d-\xi q)}=0
\EEA
The latter equation is solved by $x=(p-1-\eta)(\tilde q_d-q)
/(\eta q)$ with the same $\eta$ as in the classical case eq.(\ref{eqeta}). 
 
Both for $\omega\neq 0$ and for $\omega=0$ we multiply 
$\partial F_{quant}/ \partial\hat
q_{d\,\omega}$ by $\hat q_{d\,\omega}$ and go to the time domain. 
This yields
\bea \label{qdtau} &\beta&\mu
q_d(\tau)+M(q_d(\tau)-q_d(\tau+\epsilon))\\
&=&M\delta_{\tau,\beta} 
-\frac{\xi q^2}{(\tilde q_d-q)(\tilde q_d-\xi q)} 
\nn &+& \frac{p\beta^2 J^2}{2M}\sum_{\tau'}
q_d(\tau+\epsilon-\tau')
q_d^{p/2-1}(\epsilon-\tau')q_d^{p/2}(\epsilon+\tau') \nonumber \eea with
$\tau=j\epsilon$, $j=1,2,\cdots,M$, and $\epsilon=\beta/M$. 
Note that now eq. (\ref{muqu=}) becomes
redundant, as it follows already by summing (\ref{qdtau}) over
$\tau$. 

\subsubsection{Numerical solution at finite $M$}

We have numerically studied the quantum equations with $m=\sigma=2$.
In order to compare with the classical case where 
$(1/N)\sum S_i^2=1$, we have rescaled spins $S\to S/\sqrt{m\sigma}$,
$S^\ast\to S^\ast/\sqrt{m\sigma}$
to yield a unit constraint also in the quantum case. This
amounts to scaling  temperatures and fields 
to $T\to T/(m\sigma)^2$, $\Gamma\to \Gamma/(m\sigma)^{3/2}$.

 The Trotter limit $M\to \infty$ should be taken. 
This set of non-linear equations can be numerically solved for a
different values of $M$. We find that depending on the regularisation
term \cite{REG} we use for the static part of the free energy
eq. (\ref{Fstatic}) the low temperature behavior of the first order line shows
quite strong finite $M$ corrections and the numerical extrapolation to
the limit $M\to\infty$ is not safe. To overcome this problem we did the
following: we took two different regularisations for $F_{static}$,
i.e. we replaced the term $m\log(e^{\beta\mu}-1)$ by the general $m_1$
dependent expression,

\be
(m-m_1)\log(e^{\beta\mu}-1)+m_1
\log((1+\frac{\beta\mu}{M})^M-1)
\ee

Note that in the limit $M\to\infty$ this expression coincides with the
term $m\log(e^{\beta\mu}-1)$ for any value of $m_1$. For finite $M$ the
behavior of the first order transition line at very low temperatures
strongly depends on the value of $M$. This is a direct consequence of the
correct order of limits in the saddle point equations where the limit
$M\to\infty$ should be taken before the limit $T\to 0$. Consequently
the behavior of the line in the limit $T\to 0$ is quite different if
these limits are taken in the opposite way (i.e first, $T\to 0$ and
later on $M\to\infty$). 

In the classical model we had $M=1$,
$m_1=m$. We found that the first-order line
matches the $T=0$ axis with a negative slope without reentrance
(see figure 1). In the quantum case with $m_1=1$, $M\ge 1$ this
situation persist with smaller value for $\Gamma$ at $T=0$ and
larger slope of the transition line.
In the static approximation (eq. (\ref{Fstatic})) one has $m_1=0$,
so  the
regularisation term coincides with the $M\to\infty$ limit itself. In
this case the model shows the phenomena of reentrance for finite $M$
close to zero temperature (like in the SA of the Ising model,
 see below). As we expect the transition line to have infinite slope,
this indicates that for each $M$ an optimal value
for $m_1$ exists between $0$ and $1$ where the slope is infinite.

Numerically we proceed in the
following way: for a given value of $M$ we 
determined the value of $m_1$
such that the first order transition line meets the $T=0$ axis with
infinite slope. The {\em optimal} lines obtained in this way for
different values of $M$ (we took $M=1,3,5,7,9,11,21,41$) are then
extrapolated to $M\to\infty$. We found that a second degree polynomial
in $1/M$ is enough for such extrapolation to be accurate (even if higher
order polynomials are needed at very low temperatures). 

At zero temperature we obtain $\Gamma_c=0.5453$ for $p=4$ and 
$m_1\simeq 0.033$.  
The resulting extrapolated boundary line for $p=4$ is depicted in figure
3. Only at very low temperatures deviations form the SA are
important. The phenomena of reentrance has now dissapeared since this
was an artifact of the SA. The latent heat at very low temperatures,
across the first order transition line, vanishes exponentially with
$1/T$. At high temperatures corrections to the SA are indeed very small
and the value of the transition field in the SA is always larger than
the exact $M\to\infty$ extrapolated value. The opposite result is found
at very low temperatures.

In section III we will see that a similar scenario is
valid for Ising spins in the SA and also beyond it.

\begin{figure}
\centerline{\epsfxsize=8cm\epsffile{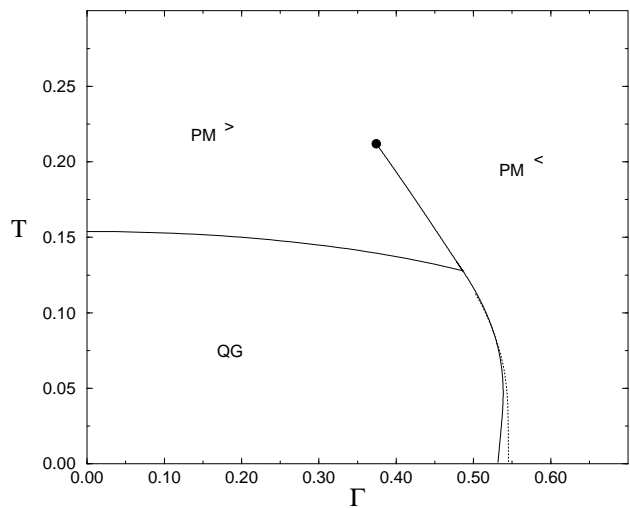}}

\caption{ Phase diagram for the quantum spherical model for $p=4$. The
multicritical point and the critical point are given in the 
static approximation (SA) by
$T_{mcp}=0.12781, \Gamma_{mcp}=0.486, T_{cep}= 0.2118,
\Gamma_{cep}=0.3743$. The continuous line is the SA and the dotted line
is the $M\to\infty$ extrapolation which yields $\Gamma_c=0.5453$ at
$T=0$.}

\end{figure}

\subsubsection{Continuum limit: $M\to\infty$}
Let us now take the limit $M\to \infty$. We set
\BEQ
q_d(\tau)=q+p(\tau)
\EEQ
with $p(\tau)=p(\tau+\beta)$. 
Eq. (\ref{qdtau}) implies a discontinuity for $\tau=0^+$:
 $p(0^+)=p(0)-1$ with $p$ being left-continuous at $0$. 
At other $\tau$ one gets
\BEA
\frac{\d p(\tau)}{\d \tau}&=&\frac{p(\tau)}{\int d\tau'\,p(\tau')}
+\frac{pJ^2}{2}\int_{-\beta/2}^{\beta/2}
\d\tau' [p(\tau)-p(\tau+\tau')]\nn
&\times& [(q+p(\tau'))^{p/2-1}(q+p(-\tau'))^{p/2}-q^{p-1}]
\EEA
Further one has
\BEQ
 q+p(0)+\frac{\Gamma^2}{\mu^2}
+\frac{m-1}{1-e^{-\beta\mu}}
= m\sigma
\EEQ
\BEQ
\frac{p(p-1)J^2}{2}q^{p-1}
=\frac{\eta q}{\large(\int\d\tau p(\tau)\large)^2}
\EEQ

\BEA
&\mu&\,=\frac{1}{\int d\tau'\,p(\tau')}+\\
\frac{pJ^2}{2}\int_{-\beta/2}^{\beta/2}
&\d&\tau' [(q+p(\tau'))^{p/2-1}(q+p(-\tau'))^{p/2}-q^{p-1}]
\nonumber
\EEA
The internal energy reads
\BEA
U&=&-\int_{-\beta/2}^{\beta/2}\d\tau 
[(q+p(\tau))^{p/2}(q+p(-\tau))^{p/2}-q^p]\nn
&-&\frac{(p-1-\eta)q^{p-1}}{\eta}
\int_{-\beta/2}^{\beta/2}\d\tau p(\tau)
-\frac{2\Gamma^2}{\mu}
\EEA
For $p=4$ its Fourier representation reads
\BEA \label{U==}
U&=&-\beta J^2\sum_{\omega_1+\omega_2=\omega_3+\omega_4}
\hat q_{d\omega_1}
\hat q_{d\omega_2}
\hat q_{d\omega_3}
\hat q_{d\omega_4}
+\beta J^2 q^4
\nn&-&	\frac{(p-1-\eta)q^{p-1}}{\eta}
\beta p_{\omega=0}
-\frac{2\Gamma^2}{\mu}\EEA
where $\hat q_{d\omega}=p_\omega+q\delta_{\omega,0}$.

These equations are particularly useful at $T=0$. It can then be seen that
$p(\tau)\sim 1/\tau^2$ for $\tau\to\pm\infty$, implying
that $p_\omega=T\int \d\tau p(\tau)e^{i\omega\tau}
\sim T(1+|\omega|+i\omega)$.
Expanding the sums in (\ref{U==}) in powers of $p$ we
find sums over 1, 2, and 3 frequencies. The one-frequency
sum can be calculated as follows. We 
 extend the Euler-Maclauren formula to complex functions
with non-analyticities of the form $|\omega|$, 
and obtain
\BEA
T\sum_{\omega=2\pi nT} f_\omega &=&\int \frac{\d\omega}{2\pi}f(\omega)
-\frac{2\pi T^2}{6}    \Re\frac{\d   f}{\d\omega  }\large\vert_{0^+}
\EEA
The two and three frequency sums are convolutions of this and
produce also $T^0+T^2$ terms.
This yields a behavior $U=U_0+U_2T^2$, which implies
a linear specific heat, $C\approx 2U_2T$, in the spin glass
phase. This result holds
for the present $p$-spin model. Unlike stated previously~\cite{Nqsg},
$C\sim T$ also holds for  the $p=2+4$ model with
infinite replica symmetry breaking.
For $p=2$ only  there is no replica symmetry breaking, 
and the system is in another universality class.
As discussed above, one then has $C\sim T^{3/2}$.

\section{Ising spins}

The Ising $p$ spin-glass model in a transverse-field is defined by,

\be
{\cal H}=-\sum_{i_1<i_2<...<i_p}
J_{i_1i_2...i_p}\s_{i_1}^z\s_{i_2}^z...\s_{i_p}^z-\Gamma\sum_i\s_i^x
\label{eq1}
\ee

where $\s_i^z,\s_i^x$ are the Pauli spin matrices and $\Gamma$ is the
the transverse field. The indices $i_1,i_2,\cdots i_p$ 
run from 1 to $N$ where $N$ is the number of sites. 
The $J_{i_1i_2\cdots i_p}$ are couplings Gaussian 
distributed with zero mean and variance $p!J^2/(2N^{p-1})$. We shall
choose units in which $J=1$.

Here we will compute in the SA the phase diagram of the model
eq.(\ref{eq1}) and show that coincides in its essentials with that
reported in the previous sections. Detailed computations of the quantum
Ising model eq.(\ref{eq1}) have been already presented in the
literature.  Here we only sketch the main steps of the derivation of the
saddle point equations skipping the details.  The interested reader will
find more details about their derivation in \cite{GOL,DOTH,CEWARAWA}.

The free energy of the model is computed using the replica method as
in the previous section. After discretizing the imaginary time
direction using the Trotter-Suzuki decomposition we obtain a problem
described by an effective Hamiltonian,

\be
{\cal H}_{eff}=A\sum_{i<j}\,J_{ij}\sum_t\s_i^t\s_j^t+B\sum_{it}
\s_i^t\s_i^{t+1} + C 
\label{eq3}
\ee

where the time index $t$ runs from 1 to $M$ and the spins $\s_i^t$ take
the values $\pm 1$. The constants $A$, $B$ and $C$ are given by
$A=\frac{\beta}{M}; B=\frac{1}{2}\ln(\coth(\frac{\beta\Gamma}{M}));
C=\frac{MN}{2}\ln(\frac{1}{2} \sinh(\frac{2\beta\Gamma}{M}))$. Now we
apply the replica trick and compute the average over the disorder of the
replicated partition function,

\be \overline{Z_J^n}=\int [dJ]
\sum_{\lbrace\s_i^t\rbrace} \, \exp(\sum_{a=1}^n\,{\cal H}_{eff}^a)
\label{eq4}
\ee Computations are easily done and the problem can be reduced
to a dynamical equation involving Ising spins in a one dimensional
chain. The free energy reads,

\be
\beta f =\lim_{n\to 0}\frac{F(Q,\La)}{n}
\label{eq4b}
\ee

where,

\bea
F(Q,\La)=-\frac{nC}{N}+\frac{1}{M^2}\Tr(Q\La)-\nonumber\\
\frac{A^2}{2}\,\sum_{abtt'}(Q_{ab}^{tt'})^p-\ln(H(\La)
)
\label{eq5}
\eea

with $Q_{ab}^{tt'},\La_{ab}^{tt'}$ being the order parameter 
and the trace $\Tr$ is done over the replica and time indices. The term
$H(\La)$ is given by,
\be
H(\La)=\sum_{\s}\,\exp(\sum_{ab}\frac{1}{M^2}\sum_{t t'}\La_{ab}^{t
t'}\s_a^t\s_b^{t'}\,+\,B\sum_{at}\s_a^t\s_a^{t+1}) 
\label{eq6}
\ee
The most general time traslation invariant solution of these equations is
given by,

\bea
Q_{ab}^{tt'}=Q_{ab}~~~(a\ne b);~~~~~~~\La_{ab}^{tt'}=\La_{ab}~~(a\ne b)\\
Q_{aa}^{tt'}=q_d(t-t')~~~~~~~~~~~\La_{aa}^{tt'}=\la_d(t-t')
\label{eq7}
\eea

Because at zero transverse field the classical solution is a one step of
replica symmetry breaking we look also for solutions of this type in the
quantum case.  We divide the $n$ replicas into $n/m$ boxes $K$ of size
$m$ such that $m$ divides $n$. The saddle point solution when $a\ne b$
takes the form $Q_{ab}^{tt'}=q; \La_{ab}^{tt'}=\la$ if $a,b\in K$ and
$Q_{ab}^{tt'}=\La_{ab}^{tt'}=0$ otherwise. If $a=b$ we have
$Q_{aa}^{tt'}=q_d(t-t'), \La_{aa}^{tt'}=\la_d(t-t')$. Finally the free
energy reads,

\bea \beta f= -C
-\frac{\beta^2}{4}(m-1)q^p-\frac{\beta^2}{4M^2}\sum_{tt'}\,(q_d(t-t'))^p
\nonumber\\+\frac{\beta^2(m-1)}{2}q\la+
\frac{\beta^2}{2M^2}\sum_{tt'}\,q_d(t-t')\la(t-t')\nonumber\\
-\frac{1}{m}\ln \bigr( \int_{-\infty}^{\infty}\d p_x\Xi^m(x)\bigl)
\label{eq8}
\eea

\noindent
and $\d p_x=\d x\exp(-x^2)/(2\pi)^{\frac{1}{2}}$ is the Gaussian measure.
The order parameters are determined by solving the saddle point equations,

\bea
\frac{\partial f}{\partial q}=\frac{\partial f}{\partial\la}=
\frac{\partial f}{\partial m}=0;\label{eq9a}\\
\frac{\partial f}{\partial q_d(t-t')}=
\frac{\partial f}{\partial\la_d(t-t')}=0;\label{eq9b}
\eea

where $\Xi(x)$ is given by,

\be
\Xi(x)=\sum_{\lbrace\s_t\rbrace}\exp(\Theta(x,\lbrace\s_t\rbrace))
\label{eq10}
\ee

with 

\bea
\Theta(x,\lbrace\s_t\rbrace))=\frac{A^2}{2}\sum_{tt'}(\la_d(t-t')-\la)
\s_t\s_{t'}\,+\nonumber\\
\,B\sum_{t}\s_t\s_{t+1}\,+A\sqrt{\la}x\sum_t\s_t~~~~~~. 
\label{eq11}
\eea
Solving equations (\ref{eq9a}),(\ref{eq9b}) we get,

\bea
\la=\frac{p}{2}q^{p-1};~~~\la_d(t-t')=\frac{p}{2}(q_d(t-t'))^{p-1};
\label{eq12a}\\
q=<<(\overline{\s_t})^2>>;~~~q_d(t-t')=<<\overline{\s_t\s_{t'}}>>
\label{eq12b}\\
\frac{(1-p)\beta^2 q^p}{4}=\frac{1}{m^2}\ln 
\bigr( \int_{-\infty}^{\infty}\d p_x\Xi^m(x)\bigl)-\nonumber\\
\frac{1}{m}<<\ln(\Xi(x))>>\label{eq12c}
\eea

where the averages $<<..>>$ and $\overline{(.)}$ are defined by,

\bea
<<A(x)>>=\frac{\int_{-\infty}^{\infty}\d p_x\Xi(x)^m\,A(x)}
{\int_{-\infty}^{\infty}dp_x\Xi(x)^m}\label{eq13a}\\
\overline{B(\lbrace\s_t\rbrace)}=\frac{\sum_{\lbrace\s_t\rbrace}
B(\s_t)\exp(\Theta(x,\lbrace\s_t\rbrace))}{\Xi(x)}
\label{eq13b}
\eea

where $\Theta(x,\lbrace\s_t\rbrace)$ is given in equation (\ref{eq11}).

The solution of this system of coupled equations is quite complex
because there is an infinity of parameters ($q_d(t-t')$) which needs to
be computed in a self-consistent way. For $p=2$ (the SK model in a
transverse field) the transition is continuous in the presence of the
transverse field and there is only one quantum paramagnetic phase. For
$p=2$ these equations have been studied using five different
methods. These are: 1) doing a self-consistent approach \cite{MIHU} or a
Ginzburg-Landau expansion \cite{RESAYE}, 2) Performing exact small $M$
calculations \cite{GOLA} 3) Perturbative expansions in the field
\cite{ISYA} , 4) Numerically solving the Schr\"odinger equation
\cite{LARI} and 5) doing quantum Monte Carlo calculations
\cite{AM,ALRI}. In the case $p\ge 3$ the transition is discontinuous and
eqs.(\ref{eq12b}),(\ref{eq12c}) have been perturbatively solved by
expanding around the $p\to\infty$ limit \cite{DOTH,CEWARAWA} where the
SA (see below) is exact. Here we will revisit the SA showing that the
phase diagram of the model coincides in its essentials with that
presented previously for the spherical model. We will go beyond the SA
later on and numerically solve
eqs.(\ref{eq12a}),(\ref{eq12b}),(\ref{eq12c}) by doing finite $M$
calculations in order to check the reliability of that approximation.

\subsection{Zeroth order solution: the static approximation}

The SA amounts to consider $q_d(t)$ and $\la_d(t)$ independent of
$t$. This corresponds to supress quantum fluctuations. This is exact at
zero transverse field but it turns out to be inaccurate at finite field
and crucial for the thermodynamic properties at zero temperature. The
failure of the SA is very clear in case of continuous quantum phase
transitions where the quantum critical point is characterized by the
dynamical exponent $z$, an exponent which cannot be computed within the
SA. The situation is slightly better in first order quantum phase
transitions where there is no critical point. Hence there is no
divergent correlation length and imaginary time correlation functions
can be well approximated by constant values \cite{COM1}. Generally this
approximation can be the source of pathologies at low temperatures where
the third principle of thermodynamics is usually violated. At not too
low temperatures we will see that this approximation yields a phase
diagram in qualitative and quantitative agreement (within $10$ per cent
in the worst case $p=3$) with the full dynamical solution.

Let us first analyze the solution of the mean-field equations in this
approximation and study the phase diagram of the model.

Introducing $\la_d(t-t')=\la_d$, $q_d(t-t')=q_d$ in the free energy
eq.(\ref{eq8}) we get 

\bea
\beta f=\frac{\beta^2(1-m)}{4}-\frac{1}{4}\beta^2
q_d^p-\frac{\beta^2(1-m)q\la}{2}-\ln(2)\nonumber\\
-\frac{1}{m}\ln\int\,\d p_x(\Xi(x))^m
\label{eq14}
\eea

where,

\bea
\Xi(x)=\int_{-\infty}^{\infty}\d p_z\cosh(T(x,z))\\
T(x,z)=(b^2+\beta^2\Gamma^2)^{\frac{1}{2}};
\label{eqT}
\eea

The $\la,\la_d,m$ are determined by solving previous equations
eqs.(\ref{eq12a}),(\ref{eq12b}),(\ref{eq12c}) and $q, q_d$ are
determined by solving the following equations \cite{COM2},

\bea
q=<<\Bigl(\overline{\sinh(T)\frac{b}{T}}\Bigr )^2>>\label{eq15a}\\
q_d=<<\overline{( \cosh(T)(\frac{b}{T})^2+\frac{\beta^2\Gamma^2 
\sinh(T)}{T^3})}>>
\label{eq15b}
\eea

where the average $<<..>>$ was previously defined in eq.(\ref{eq13a}) and
the average $\overline{(..)}$ is given by,

\be
\overline{B(x,z)}=\frac{\int_{-\infty}^{\infty}\d p_z B(x,z)}{\Xi(x)}
\label{eq15c}
\ee

The phase diagram of the model can now be computed for any arbitrary
value of $p$. As in the spherical case we find two different
paramagnetic phases. Putting $q=\la=0$ and $m=0$ in
eqs.(\ref{eq15a}),(\ref{eq15b}) they reduce to a single equation,

\be
q_d=\frac{\int_{-\infty}^{\infty}\d p_x x^2\frac{\sinh(\Phi(x))}{\Phi(x)}}
{\int_{-\infty}^{\infty}\d p_x\,\cosh(\Phi(x))}
\label{eq15}
\ee

with $\Phi(x)=\sqrt{\beta^2\Gamma^2+\beta^2\la_dx^2}$ and
$\la_d=pq_d^{p-1}/2$. This equation can be numerically solved. Like in
the spherical case one finds two paramagnetic solutions separated by a
first order transition line with latent heat. Let us call $QP^>$ and
$QP^<$ the quantum paramagnetic phases associated to the largest and
smaller value of $q_d$ respectively.  The transition line can be
constructed using the Maxwell rule. As temperature increases the latent
heat decreases. Consequently the first order line ends in a critical
point $(\beta_c,\Gamma_c)$ with mean-field critical exponents.  The
existence of this critical point has been already pointed out by
Dobrosavljevic and Thirumalai \cite{DOTH}. We have numerically computed
it for several values of $p$. Details of these computations are given in
the Appendix C. In the infinite $p$ limit this critical point is pushed
up to infinite temperature \cite{GOL} and its scaling behavior in the
large $p$ limit has been analitically obtained in \cite{DOTH} finding
$\Gamma_c=0.7579$, $T_c=0.2593\sqrt{p}$.

As temperature is lowered the first order line finishes in a
multicritical point which separates the three phases of the model (two
paramagnetic $QP^>$ and $QP^<$ and one quantum glass $QG$ ). The
boundary lines which separate the paramagnetic phases from the QG phase
correspond to different thermodynamic phase transitions. The line which
separates $QP^>$ from $QG$ has no latent heat (this is the continuation
of the usual first order classical phase transition at $\Gamma=0$).  The
line which separates $QP^<$ from $QG$ is a first order transition with
latent heat. The latent heat is positive when crossing the $QP^<\to
QP^>$ line as well as the $QP^<\to QG$ line. Lowering the temperature
the $QP^<\to QG$ line is determined by the Maxwell construction but
allowing $m$ to be different from $1$ and $q,\la$ jump to a higher
value when crossing the $QP^<\to QG$ line. For low values of $T$ the
breaking point $m$ is nearly proportional to the temperature and the
difference between $q_d$ and $q$ proportional to the temperature in the
paramagnetic (in this case $q$ is equal to 0 and $q_d$ is proportional
to $T$) as well as in the quantum glass side (where $q$ and $q_d$ reach
a finite value smaller than 1).

The behavior of the latent heat in the boundary lines $QP^<\to QP^>$ and
$QP^<\to QG$ as a function as a function of the temperature is the
following: starting from the critical point (where there is no latent
heat) and lowering the temperature the latent heat increases as a
function of the temperature reaching a maximum in the multicritical
point.  Then the latent heat decreases and vanishes like $T^{(p-1)}$ at 
low temperatures.

We have analyzed in detail the phase diagram for two different values of
$p$. We have chosen a small ($p=3$) and a large value of $p$
($p=10$). The phase diagram for $p=3$ is shown in figure 4 and that of
$p=10$ is depicted in figure 5. The latent heat along the thermodynamic
first order transition line is shown in figure 6 (for $p=3,10,\infty$
respectively). For sake of completeness we also show the dynamical
transition line for different values of $\Gamma$ in the $QP$ phase (see
\cite{RI} for more details how this line has been computed in the random
ortoghonal ($ROM$) model). The main result concerning this dynamical
line is that it crosses the first order transition $QP^>-QP^<$ below the
ending critical point. 

The main difference between figures 4,5 is that for $p=3$ the
critical point is hardly observable (but it is there!, the difference
between $T_{cp}$ and $T_{mcp}$ being of order $10^{-5}$). Also the
latent heat corresponding to $p=3$ is smaller than that of $p=10$.
Being the case $p=3$ so close to $p=2$ (where the transition is
continuous and there is no multicritical point) it is natural to find
that the transition is nearly continuous. Note that also for $p=3$ the
dynamical and the static transition lines are both very close to the
multicritical point.

\begin{figure}
\centerline{\epsfxsize=8cm\epsffile{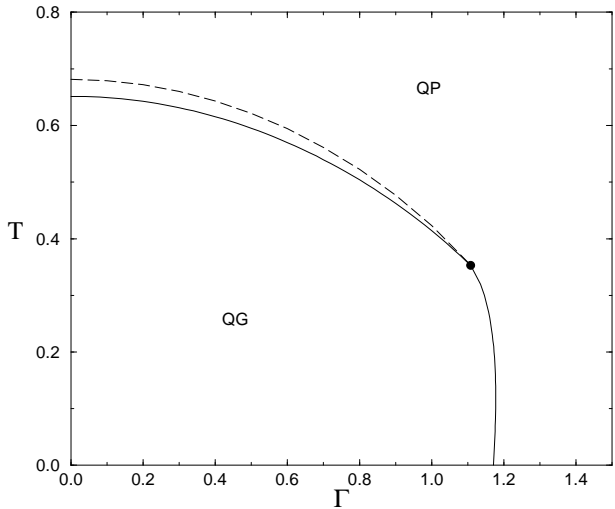}}

\caption{Phase diagram for $p=3$ with Ising spins. The critical point is
given by $T_{cp}=0.3528, \Gamma_{cp}=1.1078$. The multicritical point is
extremely close to the critical one and is indistinguishable from it in
the figure. At zero temperature, $\Gamma_c=1.174$. The dashed line is
the dynamical transition.}
\end{figure}

As anticipated in the previous sections we
observe in figure 6 that the latent heat becomes negative at very low
temperatures. For $p=3$ this happens below $T\simeq 0.1$ while for $p=10$
this effect persits but is hardly observable. This is a small
effect because the latent heat is already of order $-10^{-3}$ for
$p=3$ and $-10^{-5}$ for $p=10$. The same comments presented in the
spherical model also apply here. A negative latent heat implies
reentrace close to zero temperature. Consider the Clapeyron equation for
first order transition lines $\frac{\d\Gamma}{\d T}=L/(T\Delta M_x)$ where
$L$ is the latent heat and $\Delta M_x$ is the change in transverse
magnetisation when crossing the $PG^{<}\to QG$ line. Because $\Delta
M_x$ is always negative (increasing $\Gamma$ the transverse ordering
$M_x$ increases) a negative latent heat implies $\frac{\d\Gamma}{\d T}>0$,
i.e. reentrance.  In fact, reentrance is observed in figures 3 for $p=3$
and hardly observable (but there is) in figure $4$ for $p=10$. In the
limit $p\to\infty$ reentrance dissappears\cite{GOL}. Like in the case of
spherical quantums spins reentrance for finite $p$ is an artifact of the
SA.

Perturbing around $p=2$ we expect the following scenario to be valid:
for $p=2$ the transition is continuous (there is no latent heat) and
there is no multicritical point. Above a critical value $p_c^{(1)}\ge 2$
it appears a multicritical point which separates a first order
transition line (with latent heat) from a thermodynamic second order
transition line. The second order transition line has associated a
dynamical transition line (the dynamical transition predicted in the
framework of Mode Coupling theories) which meets the static line
precisely at the multicritical point.

\begin{figure}
\centerline{\epsfxsize=8cm\epsffile{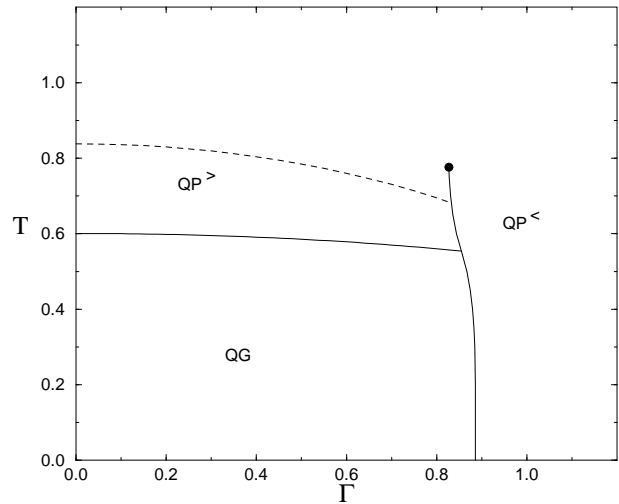}}

\caption{Phase diagram for $p=10$ with Ising spins. The critical point
and multicritical point are given by $T_{cp}=0.7765, \Gamma_{cp}=0.8903,
T_{mcp}=0.5543, \Gamma_{mcp}=0.854$. At zero temperature
$\Gamma_c=0.8855$. The dashed line is the dynamical transition.}
\end{figure}

 In the regime $2\le p\le
p_c^{(1)}$ there is a unique quantum paramagnetic phase. Above a given
value $p_c^{(2)}$ such that $p_c^{(2)}\ge p_c^{(1)}$ a first order
transition line appears with two paramagnetic phases in both
sides. Wether $p_c^{(2)}$ is larger or smaller than $3$ is
unclear. Within the SA, we expect $p_c^{(2)}$ to be quite close to
$3$. A definitive answer to this question requires a full analysis of
the theory beyond the SA. In this sense a perturbative study in
$p=2+\eps$ would be useful.  The fact that $p=3$ is close to $p_c^{(2)}$
explains why the transitions looks like a continuous one with very small
latent heat (see figure 6).

\subsection{Beyond the static approximation}

As said in the previous section it is natural to expect that the
SA works well enough if the transition is not
continuous. In fact, we expect it should yield better and better
results when $p$ increases (in the $p\to\infty$ limit it is exact)
even if it is always wrong because it definitely violates the third law of
thermodynamics at zero temperature \cite{DOTH}. For smaller values of $p$
it should be progressively worse being uncontrolled
close to the quantum transition point at $p\simeq p_c^{(2)}$.

\begin{figure}
\centerline{\epsfxsize=8cm\epsffile{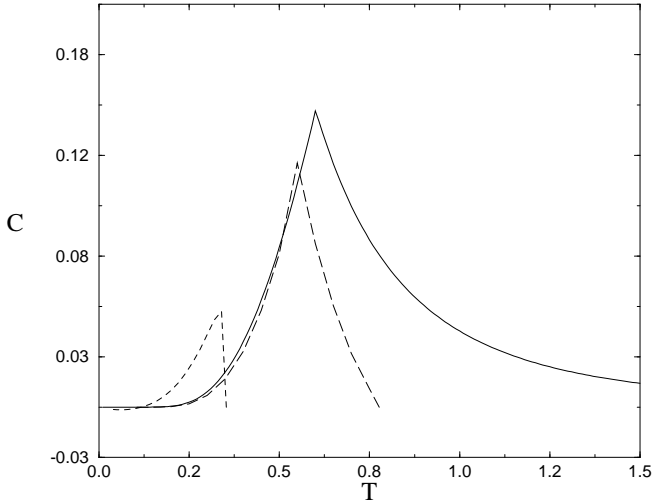}}

\caption{Latent heat for the $p$-spin Ising spin glass model with
$p=3,10,\infty$ (from left to right). The last case are the results
obtained by Goldschmidt \protect\cite{GOL}. It is shown along the
boundary lines which separate the $PM^<$ phase from the other phases as
a function of the temperature. There is a maximum at the multicritical
point.}
\end{figure}

To go beyond the SA we have numerically solved
eqs.(\ref{eq12a}),(\ref{eq12b}),(\ref{eq12c}) for different values of
$M$ for a fixed value of $\beta$ and extrapolating the results to the
$M\to\infty$ limit. This is a method which usually yields good results
and has been applied in several cases to continuous quantum phase
transtions in disordered systems \cite{GOLA,ALRI}. The essentials of the
method has been already presented in section III.F for the spherical
quantum model. Here we will show how the method works for first order
quantum phase transitions in Ising models.  Our procedure is quite
simple: we solve the system of non linear equations
(\ref{eq12a}),(\ref{eq12b}),(\ref{eq12c}) for different values of $M$
looking for a quantum paramagnetic $QP^<$ and a quantum glass $QG$
solution. We have used periodic boundary conditions such that
$\s_{M+1}=\s_1$. The $QP^<$ solution is described by $q=\la=0$ and
$q_d^{QP}(t-t')$ different from zero. Without much effort the equations
can be solved in the $QP^<$ phase up to $M\simeq 16$. In the QG phase
the solution of the set of non-linear equations requires more
computational effort (because $q,\la$ and $m$ are now finite and some
one dimensional integrals cannot be avoided). In this case we were able
to solve the equations only up to $M=12$. Looking at the crossing point
between the free energies of the two phases we can obtain the transition
point for different values of $M$. Then we extrapolate the free
energies, latent heat as well as the transition point, to the
$M\to\infty$ limit. A second degree polynomial in $1/M$ fits quite well
the data.


In figures 7 and 8 we show the free energy as a function of the
transverse field $\Gamma$ for $p=3, T=0.3$ and $p=10, T=0.4$. The
$M\to\infty$ extrapolation is compared to the static ansatz which
appears to be a reasonable approximation in this case. The error in
predicting the value of the critical field is $\simeq 10\%$ for $p=3$
($\Gamma_c^{extrap}=1.021\pm 0.002, \Gamma_c^{SA}=1.14604$) and $2\%$
for $p=10$ ($\Gamma_c^{extrap}=0.841 \pm 0.001, \Gamma_c^{SA}=0.8798$).
This error should increase at lower temperatures. The latent heat is
shown in figure 9 for different values of $M$ as well as the
extrapolation to $M\to\infty$ compared to the value obtained in the
SA. The agreement is very good for $p=10$ but not for $p=3$ where the
SA predicts a latent heat nearly 4 times larger than expected.
 
\begin{figure}
\centerline{\epsfxsize=8cm\epsffile{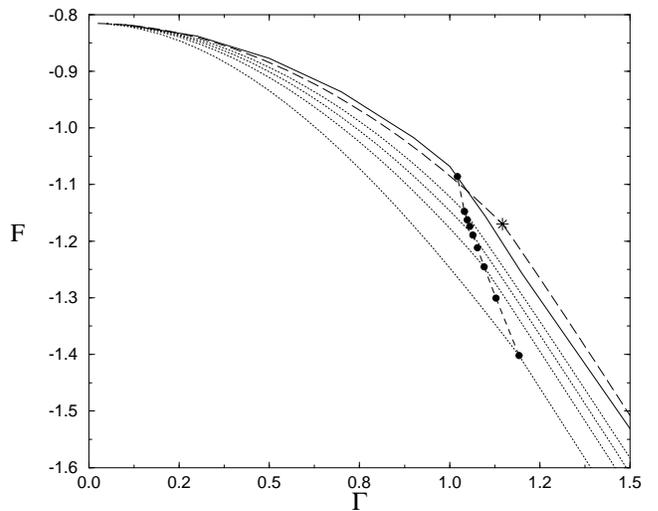}}

\caption{Free energy as a function of the transverse field $\Gamma$ in
the Ising case for $p=3$ and $T=0.3$ for different values of $M$. The
dotted lines correspond (from below to above) to $M=4,6,8,12$. The
long-dashed line is the free energy extrapolated to the $M\to\infty$
limit. The dashed line which connects the filled circles contains the
transition points for different values of $M$, the last one is the
extrapolated transition in the $M\to\infty$ limit. The continuous line
is the free energy in the SA and the star indicates the transition point
in that approximation.}

\end{figure}

\begin{figure}
\centerline{\epsfxsize=8cm\epsffile{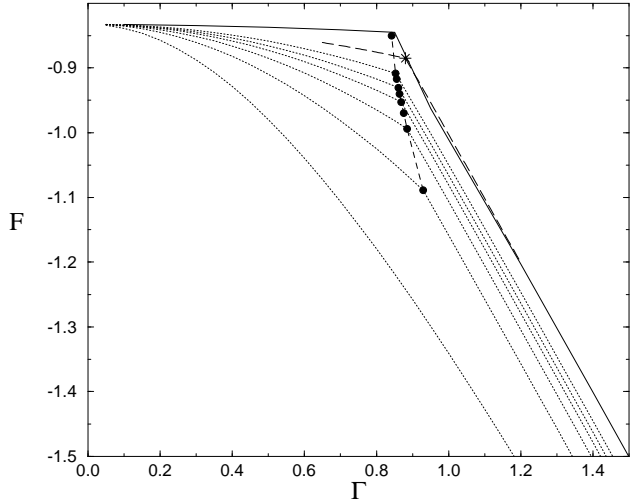}}

\caption{Free energy as a function of the transverse field $\Gamma$ in
the Ising case for $p=10$ and $T=0.4$ for different values of $M$. The
dotted lines correspond (from below to above) to $M=2,4,6,8,10,14$. The
long-dashed line is the free energy extrapolated to the $M\to\infty$
limit. The dashed line which connects the filled circles contains the
transition points for different values of $M$, the last one is the
extrapolated transition in the $M\to\infty$ limit. The continuous line
is the free energy in the SA and the star indicates the transition point
in that approximation.}

\end{figure}

Another interesting result in figures 7 and 8 concerns the jump in the
transverse magnetisation. Using the relation ${\cal M}_x=-\frac{\partial
F}{\partial \Gamma}$ this jump manifests in a discontinuous change of
the slope of the free energy as a function of $\Gamma$. From the figures
it can be observed that the transverse magnetisation always
decreases going from the $QP^<$ to the $QG$ phase. The jump is very
small for $p=3$ and increases for larger values of $p$.

It is very difficult to perform numerical calculations at much low
temperatures, mainly because the scaling behavior in $M$ is found when
the ratio $\frac{\beta}{M}$ is small in order to extrapolate to the
continuum limit $\frac{\beta}{M}\to 0$. At $T=0.1$ we have studied the
case $p=3$ for values $M=8,9,10,12,13,14$, this last case being the
limit of our computational capabilities. The results are shown in figure
9 where we plot the latent heat as a function of $\frac{1}{M}$. It is
difficult to extrapolate to $M\to\infty$ because we do not have large
enough values of $M$ in order to do that. The data is compatible with
the fact that at very low temperatures the latent heat is negligible in
the $M\to\infty$ limit.

\begin{figure}
\centerline{\epsfxsize=8cm\epsffile{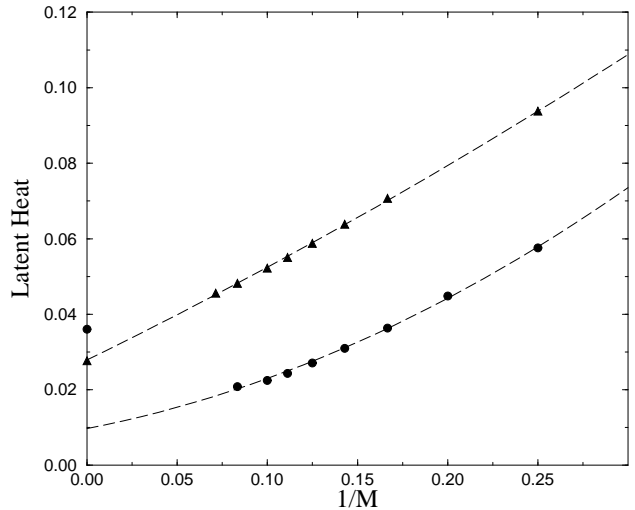}}

\caption{Latent heat for the Ising case, $p=3$ (filled circles) and
$p=10$ (filled triangles) at the transition point for $T=0.3$, $T=0.4$
respectively as a function of $1/M$. The dashed lines are second degree
polynomial fits in $1/M$ to the data. The triangle ($p=10$) and the
circle ($p=3$) in the vertical axis are the values estimated in the SA
($L^{extrap}_{p=10}=0.027\pm 0.001$, $L^{SA}_{p=10}=0.02767$,
$L^{extrap}_{p=3}=0.0097 \pm 0.001$, $L^{SA}_{p=3}=0.036$).}
\end{figure}

\section{Discussion and conclusions}

In this work we have investigated the quantum phase transition in spin
glasses with multi-spin interactions in a transverse field. We have
introduced a solvable spherical model which yields a phase diagram
qualitatively similar to that found in the Ising case (in the static
approximation -SA- and also beyond).  Details of the quantization of the
spherical model have been given. We find indications that the specific
heat is linear at low $T$. This is possibly related to a finite density
of two level systems in the free energy landscape.  We have also seen
that $p$-spin models in a transverse field (spherical and Ising)
typically have a first order transition line in the paramagnet, that we
have called the {\it pre-freezing line}.

For the $p$-spin models the study indicates (as expected) that the
static approximation (SA) can be considered as a classical approximation
where quantum fluctuations are fully neglected. A zero order calculation
shows that the SA seems to yield qualitative good results for first order
(but not too weak) phase transitions at not too low temperatures. The
situation is different for continuous quantum phase transitions.  In
particular we have checked the approximate validity of the SA in both
the spherical (with quantized spins) and the Ising model numerically
computing the free energy and the transition line. This has been done
solving the time correlator $q_d(t-t')$ for finite values of $M$ and
extrapolating to $M\to\infty$. The SA predicts the phase diagram of the
model with reasonable accuracy. For instance, for $p=3$ in the Ising
case the SA yields the phase boundaries with a precision within $10\%$
improving for larger values of $p$. The approximate validity of the SA
is restricted to high temperatures. Indeed, at very low temperatures the
SA fails. This manifests in the phase diagram of both the Ising and
spherical cases (this last one with quantized spins) which display the
phenomena of reeentrance. This pathology is related to the incorrectness
of the SA and disappears when taking into account quantum fluctuations.

\begin{figure}
\centerline{\epsfxsize=8cm\epsffile{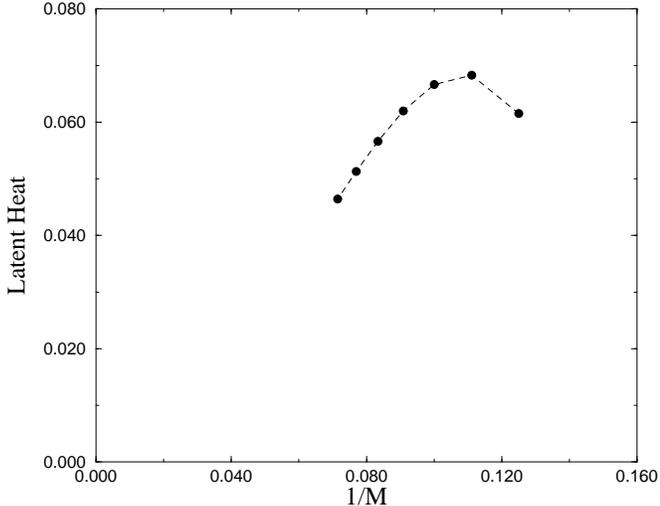}}

\caption{Latent heat for the Ising case, $p=3$ at the transition point
for $T=0.1$ as a function of $1/M$. Extrapolation to the $M\to\infty$
limit is not safe since we are far from the scaling region.}
\end{figure}

In the simplest scenario the multicritical point should appear as soon as
$1\le p_c^{(1)}<p$. In this case the phase diagram should be
qualitatively similar to that of figure 4 with only one quantum
paramagnetic phase. Above a second critical value $p_c^{(2)}$ the
multicritical point would develop a line ending in a critical point
restoring the two different quantum paramagnetic phases like is observed
in figure $5$. It would be interesting to understand (in the spherical
as well as in the Ising cases) how the phase diagram of the model
changes when expanding in $p=2+\eps$. If $2<p\le p_c^{(1)}$ the
transition should remain continuous for small $\epsilon$. Then it would be
interesting to investigate the dependence (if any) of the dynamical
exponent $z$ with $\eps$. Recent results in the $ROM$ model \cite{RI}
suggest that the quantum dynamical exponent could be not universal within
mean-field theory. This suggests that models with the same classical
behavior may display different quantum behavior in presence of the same
type of perturbation.

It  
would be very interesting to investigate the problem of the existence of
more than one quantum paramagnetic phase in the quantum Potts model
where it has been suggested (like in the $ROM$ model) that the
transition becomes continuous at zero temperature \cite{SE}. These are
subjects for future research.

\vh

{\bf Acknowledgments.} 
F.R is grateful to the Foundation for Fundamental Research of Matter (FOM)
in The Netherlands for financial support through contract number
FOM-67596. The authors acknowledge hospitality at the ISI (Turin,
Italy), where part of the work was done.

\vu
\appendix
\section{Other discretizations of the coherent state path
integral}
The coherent state path integral  has an obvious
expression in the continuum limit $M\to\infty$, $\d\tau\to 0$.
However, that is a dangerous limit, which may introduce problems that
do not occur in its finite $M$ expression~\cite{NegeleOrland}.
A typical case is the following sum over Matsubara 
frequencies
$\omega=2\pi n T$ ($n=1,\cdots,M)$
\be
P=\sum_\omega\ln(\beta \mu -i\Omega_\omega)-{\cal N}
\ee
where ${\cal N}$ is an appropiate normalization and
\be
\Omega_\omega=iMT(1-e^{i\epsilon\omega})\approx \omega 
\qquad(|\omega|\ll 1)
\ee
This sum can be carried out after expanding in powers of 
$e^{i\epsilon\omega}$ and yields in the limit $M\to \infty$
\be
P=\ln[(1+\epsilon\mu)^M-1]\to \ln[e^{\beta \mu}-1]
\ee
provided we choose ${\cal N}=M\ln M$. The common approach, however,
is to approximate $\Omega_\omega\approx\omega$, 
choose ${\cal N}=\sum_\omega (i\omega)$, and to extend to sum from
$-M/2<n\le M/2$ $\to$ $-\infty<n<\infty$, which yields the result
\be
\tilde P=\ln 2\sinh\frac{1}{2}\beta\mu=\ln[e^{\beta \mu}-1]-\beta\mu/2
\ee
This ill-defined procedure thus brings a different result
for the non-singular part. Those terms also show up in 
the zero point energy, that is to say,
terms that may arise when normal-ordering of the creation
and annihilation operators. The common approach also yields
a different answer for the
first derivative of $P$ wrt $\mu$. For the second derivative 
the convergence is quick enough to yield the same answer in both
approaches.

\section{Normalization of the path integral: free  spherical
spins in a field}\label{normalization}

The second term in eq. (\ref{a012}) is
\bea
\nn
A_1&=&(-m\sigma+\frac{m}{1-a ^M}+\frac{\Gamma^2}{\mu^2})
\sum_j\epsilon\mu_j \eea
As expected, it vanishes when $\mu$ is taken at the saddle point.
The quadratic fluctuations yield
\BEA
-2A_2&=&\epsilon^2\sum_{j}\mu_j^2(\frac{m}{1-a^M}
+\frac{\Gamma^2}{\mu^2})  \\
&+&\epsilon^2\sum_{ jj'}\mu_j\mu_{j'}(\frac{ma^M}
{(1-a^M)^2}+
\frac{\Gamma^2}{\mu^2}\frac{a^{|j-j'|}+a^{M-|j-j'|}}
{1-a^M})\nonumber
\eea
The $\mu_j^2$ and $\mu_j\mu_{j'}$ terms can be calculated by going
to Fourier space. Using the equation of motion 
 $A_2$ can be rewritten as
\bea
-2A_2&=&M\epsilon^2\sum_{\omega}\mu_\omega\mu_{-\omega}
[\sigma+\delta_{\omega,0}\frac{Mme^{\beta\mu}}{(e^{\beta\mu}-1)^2}\nn
&+&\frac{\Gamma^2}{\mu^2}
(\frac{1}{1-ae^{i\omega\epsilon}}+\frac{ae^{-i\omega\epsilon}}
{1-ae^{-i\omega\epsilon}})]
\eea
Thus the $\mu$-integrals yield
\bea
Z&=&\frac{C_M}{(2\pi MN\sigma)^{M/2}}
\sqrt{\frac{\sigma+\frac{2M\gamma}{\beta\mu}}
{\sigma+\frac{Mme^{\beta\mu}}{(e^{\beta\mu}-1)^2}+
\frac{2M\gamma}{\beta\mu}}}e^{-A_0-\frac{1}{2}D}
\eea
\bea
D=\sum_\omega
\ln{\frac{1+a^2+\frac{\gamma}{\sigma}(1-a^2)-ae^{i\omega\epsilon}
-ae^{-i\omega\epsilon}}
{(1-ae^{i\omega\epsilon})(1-ae^{-i\omega\epsilon})}}
\eea
The $\omega$-sums can be carried out using 
\be \sum_\omega\ln(1-be^{\pm i\omega\epsilon})=\ln(1-b^M)
\ee 
For the leading behavior at $\Gamma\neq 0$ we get 
with $b=1-\Gamma\sqrt{2\epsilon/\mu} $
\be 
D\approx M\ln\frac{1}{b}+2\ln(1-b^M)\approx
\sqrt{M}\sqrt{\frac{2\beta\Gamma^2}{\mu}}
\ee
If we choose 
\be C_M=({2\pi MN\sigma})^{M/2}\ee
 it follows for $\Gamma\neq 0$
that the free energy has, on top of the extensive part $TA_0$,
a non-universal contribution of order $N^0M^{1/2}$.
For $\Gamma$ stricktly equal to 0, there is a universal term
$N^0\ln M$. Both terms are non-extensive and can be omitted if one 
first takes $N$ large and then $M$~\cite{Nqsm}. Actually this is also
the limit that underlies the saddle point approximation. 
Physically it is also the natural limit, 
as  for fixed small $T$ the $M=\infty$ limit is reached already 
for $M\sim 1/T$ independent of $N$. This example shows that
the extensive part of the free energy of quantum spherical spins
is a well defined, natural object. Non-extensive parts are more
delicate.

\section{Critical endpoint in the static approximation: Ising case}

\vh

Here we give the equations which yields the critical endpoint in the SA
in the Ising case (see also \cite{DOTH} for the original derivation).

Starting from eq.(\ref{eq15}) we define the function $g(x)$,

\be
g(q)=q-\overline{
\Phi(x)^{-1}x^2\sinh\Phi(x)}
\label{A1}
\ee

where $\Phi(x)=\sqrt{\beta^2\Gamma^2+\beta^2\la x^2}$ and
$\la=pq^{p-1}/2$ and the average
$\overline{(.)}$ is defined by,

\be
\overline{A(x)}=\frac{\int_{-\infty}^{\infty}\d p_x
A(x)}{\int_{-\infty}^{\infty}\d p_x \cosh(x)}
\label{A2}
\ee

The paramagnetic phases are found by solving the equation
$g(q_d)=0$. This yields one solution at very high temperatures and three
solutions at lower temperatures. Of these three solutions two of them
are stable (the ones with largest and smallest values of $q_d$) while
the other one (that with an intermediate value of $q_d$) is
unstable. This is the same scenario as in the spherical model. The
critical point is then determined by the coalescence of these two stable
solutions. This gives the equations,

\be
g(q_d)=\bigl (\frac{\partial g}{\partial q}\bigr )_{q=q_d}=
\bigl (\frac{\partial^2 g}{\partial q^2}\bigr )_{q=q_d}=0
\label{A3}
\ee

These three equations read,

\bea
q_d=f_c^{(1)}(q_d)=q_d\\
\frac{\beta^2 p(p-1)}{4} q_d^{p-2}\,f_c^{(2)}(q_d)=1\\
(2-p)\,f_c^{(2)}(q_d)=\frac{\beta^2 p (p-1) q_d^{p-1}}{4} f_c^{(3)}(q_d)
\eea

where $f_c^{(1)}, f_c^{(2)},f_c^{(3)}$ are the first three 
cumulants associated to the functions $f^{(n)}$
($n=1,2,3$),

\bea
f^{(1)}&=&\overline{
\Phi^{-1}x^2\sinh\Phi }
\\
f^{(2)}&=&\overline{
\Phi^{-3}x^4(\Phi\cosh\Phi-\sinh\Phi)}
\\
f^{(3)}&=&\overline{
\Phi^{-5}x^6(\Phi^2\sinh\Phi
-3\Phi\cosh\Phi+3\sinh\Phi)}
\eea

These equations can be exactly solved yielding $T_{cp},\Gamma_{cp},q_d^{cp}$
for different values of $p$. Is not difficult to generalize this set of
equations beyond the SA in the general case.

\vu

\section{Equations for the energy in the Ising case}

In this appendix we give the exact expressions for the internal energy
used to compute the latent heat in section IV.B. We start from equation 
(\ref{eq8}) by evaluating the derivative $u=\frac{\partial \beta
f}{\partial \beta}$. This yields,

\bea
u=-\frac{\partial C}{\partial\beta}-\frac{\beta (m-1)
q^p}{2}-\frac{\beta}{2M^2}\sum_{tt'}(q_d(t-t'))^p\,+\nonumber\\
\beta (m-1)
q\la\,+\,\frac{\beta}{M^2}\sum_{tt'}q_d(t-t')\la(t-t')\nonumber\\
-\frac{1}{m}\frac{\int_{-\infty}^{\infty}dp_x\Xi^{m-1}(x)\frac{\partial
\Xi}{\partial\beta}}{\int_{-\infty}^{\infty}dp_x\Xi^m(x)}
\label{B1}
\eea

where $C$ and $\Xi(x)$ were defined in eqs.(\ref{eq3}) and (\ref{eq10})
respectively and $dp_x$ is the Gaussian measure. Doing the last
integral by parts and rearranging terms we get the final expression,

\bea
u=-\Gamma\coth(\frac{2\beta\Gamma}{M})-\frac{\beta(m-1)q^p}{2}-\nonumber\\
\frac{\beta}{2M^2}\sum_{tt'}(q_d(t-t'))^p+\frac{\Gamma}
{\sinh(\frac{2\beta\Gamma}{M})}\,q_d(1)
\label{B2}
\eea

In the continuum limit $M\to\infty$ the $q_d(t)$ becomes a continuous
function of time yielding,

\be
u=\Bigl (\frac{\partial q_d}{\partial t}\Bigr )_{t=0}
-\frac{\beta (m-1) q^p}{2}-\frac{1}{2}\int_0^{\beta} (q_d(t))^pdt
\label{B3}
\ee

In the $QP^<$ phase at zero temperature in the large $\Gamma$ regime 
we have $q_d(t)\sim \exp(-t\Gamma)$ yielding $u\sim
-\Gamma-1/(2p\Gamma)$. Note that the SA is only exact in the limit 
$p\to\infty$ where the energy is given by $u=-\Gamma$.

It is also easy to check that in the SA the energy is simply given by,

\be
u=-\frac{\beta}{2}(q_d^p-(1-m)q^p)-\beta\Gamma^2
<<\overline{\frac{\sinh(T(x,z))}{T(x,z)}}>>
\label{B4}
\ee

where the averages $\overline{(..)}$, $<<(..)>>$ and $T(x,z)$ where
previously defined in eqs.(\ref{eq13a}), (\ref{eq15c}), (\ref{eqT})
respectively.


\begin{thebibliography}{99}

\bi{QUANTUM_REVIEWS} S.~Sachdev in {\em Statphys 19}, Ed. Hao Bailin,
(World Scientific, Singapore 1996); H.~Rieger and A.P.~Young, Review
article for XIV Sitges Conference: Complex Behavior of Glassy
Systems. Lecture Notes in Physics, Ed. by M. Rub\'{\i}, Springer Verlag,
Heidelberg (1997) in press, preprint {\bf cond-mat/9707005};
S.L.~Sondhi, S.M.~Girvin, J.P.~Carini and D.~Shahar,
Rev. Mod. Phys. {\bf 69}, 315 (1997).

\bi{BOOK} B. K. Chakrabarty, A. Dutta and P. Sen, {\em Quantum Ising
Phases and Transitions in Transverse Ising models}, Lecture Notes in
Physics {\bf M41}, Springer-Verlag, Berlin 1996.

\bi{YOSA} S. Sachdev and A. P. Young, Phys. Rev. Lett. {\bf 78} 2220 (1997)

\bi{BERNA1} J. P. Bouchaud and M. Mezard, J. Physique I (Paris) {\bf 4}
(1994) 1109; 

\bi{BERNA2} E. Marinari, G. Parisi and F. Ritort, J. Phys. A
(Math. Gen.)  {\bf 27} 7615 (1994); 

\bi{ROM} E. Marinari, G. Parisi and F. Ritort, J. Phys. A (Math. Gen.)
{\bf 27} 7647 (1994);

\bi{BOOKS} K.~Binder and A.P.~Young, Rev. Mod. Phys. {\bf 58}, 801
(1986); M.~M\'ezard, G.~Parisi and M.A.~Virasoro, {\em Spin Glass
Theory and Beyond} (World Scientific, Singapore 1987); K.H.~Fischer and
J.A.~Hertz, {\em Spin Glasses} (Cambridge University Press, Cambridge 1991).

\bi{KITHWO} T. R. Kirkpatrick and D. Thirumalai, Phys.  Rev. {\bf B36}
(1987) 5388 ; T. R. Kirkpatrick and P. G. Wolynes, Phys. Rev. {\bf B36}
(1987) 8552;

\bi{THEO1} Th. M. Nieuwenhuizen. {\em Complexity as the driving force for
glassy transitions} Preprint {\bf cond-mat 9701044}

\bi{AGM} J. H. Gibbs and E. A. Di Marzio, J. Chem. Phys. {\bf 28} 373
(1958); G. Adams and J. H. Gibbs, J. Chem. Phys. {\bf 43} 139 (1965);

\bi{ALFRRI} D. Alvarez, S. Franz and F. Ritort, Phys. Rev {\bf B54} (1996)
9756

\bi{NThermo} Th.M. Nieuwenhuizen, preprint (1996)

\bi{COM1} We suppose that the size of the system is infinite from the
begining or that the typical time scale involved in the adiabatic
process grows with the size of the system much slower than any
exponentially long time scale.

\bi{RI} F.~Ritort, Phys. Rev. {\bf B55}, 14096 (1997); see also {\em
Classical and Quantum Behavior in Mean-Field Glassy Systems} Review
article for XIV Sitges Conference: Complex Behavior of Glassy
Systems. Lecture Notes in Physics, Ed. by M. Rub\'{\i}, Springer Verlag,
Heidelberg (1997) in press, preprint {\bf cond-mat/9611034};


\bi{BM} A.J.~Bray and M.A.~Moore, {\sl J. Phys C.} {\bf 13}, L655 (1980).

\bi{CRSO} A. Crisanti and H. J. Sommers, Z. Phys. {\bf B92} (1992) 341;
A. Crisanti, H. Horner and H.-J. Sommers, Z. Phys. {\bf B92} (1993) 257;

\bibitem{p=3cr} In the case $p=3$, $m\sigma=1$, $ m=2$
one has $(T_{cep},\Gamma_{cep})$ $=(0.4070,\, 0.89001)J$;
($T_{mcp},\Gamma_{mcp})$ $=(0.37506,\, 0.95379)J$;

\bi{PSW} R. Pandit, M. Schick, and M. Wortis, Phys. Rev. B {\bf 26} 
(1982) 5112

\bi{FLN} G. Forgacs, R. Lipowsky, and Th.M. Nieuwenhuizen, in:
{\it Phase Transitions and Critical Phenomena }{\bf 14},
C. Domb and J.L. Lebowitz, eds, pp 135 (Academic, New York, 1991)

\bi{RutTab} J.E. Rutledge and P. Taborek,
Phys. Rev. Lett. {\bf 69} (1992) 937

\bi{Bonn} D. Bonn, H. Kellay, and G.H. Wegdam, J. Phys. Cond. Mat. 
{\bf 6} (1994) A389

\bi{Kellay} H. Kellay, D. Bonn, and J.M. Meunier,
Phys. Rev. Lett. {\bf 71} (1994) 2607

\bi{GAR} D. J. Gross and M. Mezard, Nucl. Phys. {\bf B240}
(1984) 431; E. Gardner, Nucl. Phys.  {\bf B257} (1985) 747.

\bi{REM} B. Derrida, Phys. Rev. {\bf B24}  2613 (1981).

\bi{GOL} Y. Y. Goldschmidt, Phys. Rev. B {\bf 41}, 4858 (1990);

\bi{MOT} P. Mottishaw, Europhys. Lett. {\bf 1}, 409 (1986)

\bi{DOTH} V. Dobrosavljevic and D. Thirumalai, J. Phys. A {\bf 22}, L767
(1990); 
  
\bi{CEWARAWA} L. De Cesare, K. Lubierska-Walasek, I. Rabuffo and
K. Walasek, J. Phys. A {\bf 29}, 1605 (1996);

\bi{FRPA} S. Franz and G. Parisi, {\em Phase diagram of glassy systems
in an external field} Preprint {\bf cond-mat 9701033}

\bi{Nqsg} Th. M. Nieuwenhuizen, Phys. Rev. Lett. {\bf 74} (1995) 4289;

\bi{Nqsm} Th. M. Nieuwenhuizen, Phys. Rev. Lett. {\bf 74} (1995) 4293;

\bi{NegeleOrland} J.W.  Negele and H. Orland, 
{\it Quantum Many-Particle Systems},
Addison-Wesly, Redwood City (1988).


\bi{Hart-Weichman}
J.W. Hartman and P.B. Weichman, Phys. Rev. Lett {\bf 74} (1995)
4584

\bi{Voita} T. Voita, Phys. Rev. B {\bf 53} (1996) 710

\bi{VoitaSchreiber} T. Voita and M. Schreiber, Phys. Rev B {\bf 53}
(1996) 8211 

\bi{KTJ} J.M. Kosterlitz, D.J. Thouless, and R.C. Jones,
Phys. Rev. Lett. {\bf 36} 1217 (1976)

\bi{N85} Th.M. Nieuwenhuizen, Phys. Rev. B {\bf 31} R7487 (1985)

\bi{REG} Regularisation terms correspond to $M$ dependent expressions 
which yield the same $M\to\infty$ limit.

\bi{MIHU} J. Miller and D. Huse,
Phys. Rev. Lett. {\bf 70}, 3147 (1993).

\bi{RESAYE} J. Ye, S. Sachdev
and N.~Read, Phys. Rev. Lett. {\bf 70}, 4011 (1993); N. Read  and
S. Sachdev, Phys. Rev {\bf B52} 384.

\bi{GOLA} G.B\"{u}ttner and K.D.~Usadel, Phys. Rev. B {\bf 41}, 428
(1990); Y.Y.~Goldschmidt and P.Y.~Lai, Phys. Rev. Lett. {\bf 64}, 2467 (1990).

\bi{ISYA} H.~Ishii and Y.~Yamamoto, J. Phys. C {\bf 18}, 6225 (1985); 
J. Phys. C {\bf 20}, 6053 (1987).

\bi{LARI} D. Lancaster and F. Ritort, J. Phys. A {\bf 30} L41
(1997). For a recent investigation see also, P. Sen, P. Ray and
B. K. Chakrabarti, preprint cond-mat/ {\bf 9705297}.

\bi{AM} For a numerical study in two and three dimensions see, M. Guo,
R. N. Bhatt and D. A. Huse, Phys. Rev. Lett. {\bf 72}, 4137 (1994);
H. Rieger and A. P. Young, Phys. Rev. Lett. {\bf 72}, 4141 (1994).

\bi{ALRI} J.V.~Alvarez and F.~Ritort, {\sl J. Phys. A} {\bf 29} (1996) 7355.

\bi{COM1} Obviously, for weak first order phase transition this
assertion is no longer true.

\bi{COM2} Note that the static ansatz, in the general case,
cannot be in agreement with eq.(\ref{eq12b}) where the one dimensional
structure of the Hamiltonian $\Theta(x,\lbrace\s_t\rbrace)$ prevents
the $q_d(t-t')$ to be time independent.

\bi{SE} T.~Senhil and S.N.~Majumdar, Phys. Rev. Lett. {\bf 76}, 3001 (1996).



 \end{thebibliography}
\end{document}